\documentclass[11pt,tightenlines,eqsecnum,floats,aps,amsmath,amssymb,nofootinbib,prd,shownopacs,floatfix]{revtex4}

\usepackage{amsmath,amssymb,amsfonts,dcolumn,color,graphicx,graphics,latexsym}
\usepackage[mathscr]{eucal}
\usepackage[latin1]{inputenc}
\usepackage{enumerate}
\newcommand{\be}{\begin{equation}}
\newcommand{\ee}{\end{equation}}
\newcommand{\ba}{\begin{eqnarray}}
\newcommand{\ea}{\end{eqnarray}}

\newcommand{\lp}{\mathrm{\ell}_{\mathrm{Pl}}}

\newcommand{\bmult}{\nopagebreak[3]\begin{multline}}
\newcommand{\emult}{\end{multline}}

\def\d{{\rm d}}

\begin{document}
	
\title{Von Neumann stability of modified loop quantum cosmologies}
\author{Sahil Saini}
\email{ssaini3@lsu.edu}
\author{Parampreet Singh}
\email{psingh@lsu.edu}
\affiliation{ Department of Physics and Astronomy,\\
Louisiana State University, Baton Rouge, LA 70803, U.S.A.}

\begin{abstract}
Von Neumann stability analysis of quantum difference equations in loop quantized spacetimes has often proved useful to understand viability of quantizations and whether general relativistic  description is recovered at small spacetime curvatures. 
We use this technique to analyze the infra-red behavior of quantum Hamiltonian constraint in recently explored modifications of loop quantum cosmology: mLQC-I and mLQC-II, for the spatially flat FLRW model. We investigate the behavior for $\mu_o$ scheme, where minimum area of loops in quantization procedure does not take physical metric in to account, and the $\bar \mu$ scheme where quantization procedure uses physical metric. The fate of stability of quantum difference equations is tested for massless scalar field as well as with inclusion of a positive cosmological constant. We show that for mLQC-I and mLQC-II, difference equation fails to be von Neumann stable for the $\mu_o$ scheme if cosmological constant is included signaling problematic behavior at large volumes. Both of the modified loop quantum cosmologies are von Neumann stable for $\bar \mu$ scheme. In contrast to standard loop quantum cosmology, properties of roots turn out to be richer and intricate. Our results demonstrate the robustness of $\bar \mu$ scheme (or `improved dynamics') in loop quantization of cosmological spacetimes even when non-trivial quantization ambiguities of Hamiltonian are considered, and show that $\mu_o$ scheme is non-viable in this setting.
\end{abstract}
	
\maketitle

\section{Introduction}

Loop quantum cosmology (LQC) is an application of techniques of loop quantum gravity (LQG) to quantize symmetry reduced spacetimes, which leads to a quantum description where the quantum Hamiltonian constraint is a difference equation instead of a differential equation as in Wheeler-DeWitt theory \cite{as-status}. The origin of difference equation is tied to the underlying quantum geometry which also fixes the discreteness in the quantum difference equation.  The high level of symmetry reduction performed at the classical level in LQC often leads to various routes to quantization leading to quantization ambiguities in these models which may even differ qualitatively in their predictions. An important exercise is to rule out some of these ambiguities by  performing various physical consistency checks, such as checking the predictions of the quantized model in large volume and low curvature limit where it should agree with classical general relativity (GR). One way to ensure the correct infra-red behavior, or agreement with GR at large volumes, is to demand that the solutions of the  difference equations approximate the solutions of the Wheeler-DeWitt quantization in the infra-red limit which is known to capture solutions of GR extremely well at macroscopic scales (see for eg. Refs. \cite{aps2,aps3}). This problem can be studied by using von Neumann stability analysis, which is conventionally used to understand viable discretizations of a given differential equation. In the present context, where quantum difference equations are fundamental, it has proved to be useful tool to gain insights on understanding infra-red behavior of quantum Hamiltonian constraint in cosmological \cite{Bojowald2003,date2,Cartin2005,Cartin2005a,Rosen2006,martin-cartin-khanna,Nelson2009,tanaka,khanna-review,ps12} and black hole spacetimes \cite{Cartin2006,Yonika2018}. Interestingly, this technique provides a valuable complementary path to phenomenological studies using Friedmann and Raychaudhuri equations which are often used to rule out various quantization ambiguities by demanding a consistent infra-red behavior \cite{cs-unique}. Quite often von Neumann instability is linked to phenomenological problems discovered independently using effective Hamiltonian dynamics, and von Neumann stability guarantees viable physical description at macroscopic scales \cite{ps12}.   Our goal in this manuscript is to apply the von Neumann stability analysis to two quantizations of the spatially flat FLRW model first proposed in \cite{Yang2009}, and  revived in recent studies \cite{Dapor2017,tomasz,Li2018,ps_mLQC-II}, which differ in non-trivial ways from LQC.  

The standard loop quantization of the spatially flat FLRW model carried out in \cite{Ashtekar2003,aps1,aps2,aps3} utilized the fact that after symmetry reduction, the two terms constituting the Hamiltonian constraint (referred to in the literature as the Euclidean and the Lorentzian terms), when expressed in terms of the holonomy-flux variables of LQG, turn out to be proportional to each other. Thus, the Lorentzian term can be subsumed into the Euclidean term before quantization, leading to `standard LQC'. But one can treat Euclidean and Lorentzian terms separately on lines of LQG using Thiemann's regularization of Hamiltonian constraint \cite{thiemann}, which is generally the route since combining these two terms is only possible when certain symmetries are present. This led Yang, Ding and Ma to propose two modifications of LQC \cite{Yang2009} (or mLQC for modified LQC \cite{ps_mLQC-II}) which have been recently found to yield qualitatively different physics than LQC  \cite{Dapor2017,tomasz,Li2018,ps_mLQC-II} while preserving main features of generic resolution of cosmological singularities \cite{saini-singh-mlqc}. The two modifications of LQC are referred to as mLQC-I and mLQC-II (following notation of Ref. \cite{Li2018}). The mLQC-I quantization proceeds by treating Euclidean and Lorentzian terms independently, where as mLQC-II is obtained by further utilizing the fact that the Ashtekar-Barbero connection in case of spatially flat FLRW model is proportional to the extrinsic curvature of spatial slices. Thus, the extrinsic curvature in the Lorentzian term can be substituted by the Ashtekar-Barbero connection, leading to a qualitatively different quantum Hamiltonian constraint which results in mLQC-II. However, it should be noted that mLQC-II can arise only in a  spatially flat spacetime, such as spatially flat FLRW model or a spatially flat Bianchi-I spacetime. In contrast, mLQC-I which does not require equating Ashtekar-Barbero connection with extrinsic curvature captures the general feature of quantization of various spacetimes in LQG. 

Here let us note that one of the fundamental problems of understanding cosmological spacetimes in LQC is the way LQC fits in to LQG. In this respect, mLQC quantizations, especially mLQC-I promises to yield important new insights. It is to be emphasized that mLQC-I arises independently from attempts to obtain a cosmological sector of LQG using complexifier coherent states \cite{Dapor2017}, and mLQC-II can be understood as a possible variant of mLQC-I when spatial flat spacetimes are considered. Though, there can be in principle many modifications of LQC in the Planck regime, it is to be noted that till date only mLQC-I and mLQC-II are the ones where a physical Hilbert space is available and action of quantum Hamiltonian constraint along with effective dynamics has been studied in detail. At present, these are only two well motivated alternatives to LQC. In addition, mLQC-I and mLQC-II are closer to the full LQG because of the way Euclidean and Lorentzian terms in the quantum Hamiltonian constraint are treated. Note that treating these terms independently is expected to be a generic feature of spacetimes in LQG, not just cosmological spacetimes. 
Thus, studying properties of the quantum Hamiltonian constraints in mLQC-I and mLQC-II is not only an important exercise in itself but is also useful to understand some of the general features of the quantizations of such spacetimes in LQG.   
Further, as discussed below, the quantum hamiltonian constraint in mLQC-I and mLQC-II turns out to be a fourth order difference equation in contrast to a second order difference equation in LQC. So far little is understood on the von Neumann stability properties of fourth order difference equations in loop quantization of cosmologies. Since such higher order difference equations are expected to be more general in existence than the second order difference equation as in standard LQC,  understanding von Neumann stability of quantum Hamiltonian constraint in mLQC-I and mLQC-II, which directly tells us about the large volume behavior of these quantizations, becomes a pertinent issue. One may expect that such an investigation will shed insights on the von Neumann stability of similar quantizations with higher order difference equations. 

The mLQC quantizations are different from standard LQC in many ways. The difference equations of mLQC are fourth order compared to the second order difference equation obtained in standard LQC. The effective dynamics of mLQC-I leads to an asymmetric bounce where the present classical universe is connected through the bounce to a past universe that remains quantum in the asymptotic past \cite{Dapor2017,Li2018}.
This is in contrast to standard LQC where the universe is classical both in asymptotic past and asymptotic future but passes through a Planckian quantum phase only at the bounce. In contrast to mLQC-I, even after non-trivial changes in the structure of quantum Hamiltonian constraint the effective dynamics of mLQC-II leads to a symmetric bounce qualitatively similar to standard LQC \cite{Yang2009,Li2018}.  Due to the fact that the difference equations are now fourth order, these models lead to a much richer phenomenology compared to standard LQC. Recall that in Planck regime, modified Friedmann dynamics in LQC results in a bounce via modifications which are quadratic in energy density \cite{aps3}. Whereas, the common feature of mLQC-I and mLQC-II is that they lead to modifications in effective Friedmann and Raychaudhuri equations with terms of order higher than quadratic in energy density \cite{Li2018,ps_mLQC-II}. The kinematical phase space of the effective dynamics is shown to consist of two different branches owing to different roots of the conjugate variable ($ b $) of volume, rather than just one branch in LQC. In mLQC-I, the two branches are intertwined with each other at the bounce to describe the evolution.
Whereas in mLQC-II, only one branch describes the entire evolution  while the other branch is unphysical and irrelevant. The emergence of these new quantizations calls for a thorough investigation of the properties of quantum difference equations and their infra-red behavior to gain insights on their viability.  

Let us recall that there are additional quantization ambiguities arising from the choice of loops for evaluating holonomies of the connection in the field strength in the Hamiltonian constraint. These are named the $ \mu_o $ quantization (or `old LQC') \cite{Ashtekar2003,aps2} and the $ \bar{\mu} $ quantization or `improved dynamics' introduced in Ref. \cite{aps3}. The key difference in these quantizations is that the $\mu_o$ scheme does not use the information about the physical metric in assignment of areas of the loops, resulting in $\mu_o$ to be a numerical constant. In contrast, $\bar \mu$ scheme takes physical metric carefully in to account. In the latter, $\bar \mu$ is not constant but depends inversely on the scale factor in the physical metric. It is well known that the $\mu_o$ quantization in standard LQC is problematic due to several reasons \cite{aps3,as-status}, including lack of an infra-red limit which becomes clearly evident for matter violating strong energy condition \cite{cs-unique,cs-geom,noui}. As an example, in presence of a positive cosmological constant the $\mu_o$ quantization of standard LQC results in large deviations from GR at late times leading to an unphysical recollapse \cite{cs-unique}. This deficiency, first seen via effective dynamics, is captured by von Neumann instability of the quantum difference equation in standard LQC \cite{tanaka,ps12}. Interestingly, for matter which satisfies strong energy condition, such as a massless scalar field, such effects are tamed and LQC difference equation is von Neumann stable. But since $\mu_o$ scheme fails for inflationary spacetimes or in presence of positive cosmological constant it is phenomenologically non-viable. In contrast, all the problems of $\mu_o$ scheme were overcome in $\bar \mu$ scheme in standard LQC. In this case, the quantum difference equation is von Neumann stable for matter ranging from a massless scalar field to a positive cosmological constant, thus covering the entire range of weak energy condition, resulting in a  viable phenomenology (see Ref. \cite{agullo-singh} for a review). The ambiguities of $\mu_o$ and $\bar \mu$ schemes are also present in mLQC-I and mLQC-II.  In fact, the first mLQC-I quantization of the spatially flat FLRW model was  obtained in the $ \mu_o $ scheme in \cite{Bojowald2002}. Later, both mLQC-I and mLQC-II quantizations using the $ \bar{\mu} $ scheme were obtained in \cite{Yang2009}. The work of Dapor and Liegener \cite{Dapor2017} on obtaining an effective Hamiltonian using coherent states for mLQC-I is also based on $\mu_o$ scheme. Using effective dynamics, phenomenological implications of $\bar \mu$ scheme for inflationary spacetimes have been studied in Ref. \cite{ps_mLQC-II} for mLQC-I and mLQC-II, but little is known about the viability of $\mu_o$ scheme in mLQC-I and mLQC-II.

It is to be noted that the application of von Neumann stability analysis to the difference equations of loop quantum cosmologies differs in motivation from its usual applications in numerical methods.  In its conventional applications, von Neumann stability analysis is applied to analyze the efficacy of finite difference schemes used to approximate partial differential equations (PDEs).  Stability ensures that the solution of the finite difference scheme remains close to the solution of the PDE.   In contrast, in loop quantization of cosmological spacetimes there is no fundamental differential equation that the difference equation  approximates. Rather the quantum difference equation obtained in the process of quantization is fundamental.  However, we expect the semi-classical solutions of these difference equations to approximate GR very well in the low curvature and large volume limit, which can be ensured by demanding that these solutions approximate the solutions of (differential) Wheeler-DeWitt equation. However, these solutions of difference equation  must be stable under  fluctuations in approximating GR in the large volume limit, i.e. these fluctuations  must not grow unbounded in loop quantum evolution as universe becomes large and classical. It is in this context that we apply the von Neumann stability analysis to loop quantum difference  equations in the large volume limit to ensure the correct infra-red behavior. As we will see, the von Neumann stability analysis leads to a condition on the roots of an algebraic equation obtained from Fourier analysis of the difference equations which yields important information on the stability or instability of the underlying difference equation. 

The plan of this manuscript is as follows. We begin in Sec. IIA by reviewing different routes to loop quantization of spatially flat FLRW model that lead to quantization ambiguities in Hamiltonian constraint. For completeness, we discuss stability of difference equations in Sec. IIB and describe the von Neumann stability analysis. In Sec. III, we analyze the stability of the mLQC-I model using both $ \mu_o $ and $ \bar{\mu} $ schemes, for massless scalar field as well as positive cosmological constant,   and, in Sec. IV we analyze the mLQC-II model in both the schemes.  We find that though mLQC-I and mLQC-II yield von Neumann stable difference equations for $\mu_o$ scheme in the case of massless scalar field, the difference equations are unstable when cosmological constant is included. This signals problematic behavior at large volumes as is the case with standard LQC.  However, the difference equations in $ \bar{\mu} $ scheme turn out to be stable in both mLQC-I and mLQC-II models. We conclude with a summary of results in Sec. V.
	
\section{Preliminaries} \label{preliminaries}

In Sec. \ref{quantizations}, we lay out the procedure to obtain Hamiltonian constraint for the spatially flat FLRW model in terms of Ashtekar variables in LQC. This allows us to identify some of the quantization ambiguities in the process which result in modified loop quantum cosmologies (mLQC-I and mLQC-II). As mentioned in the introduction, unlike LQC, in mLQC-I and mLQC-II the Lorentzian and the Euclidean terms are quantized separately without combining them at the classical level. Differences between mLQC-I and mLQC-II arise from the treatment of extrinsic curvature as a connection in the latter. In this section, we  also discuss the physical differences between the $ \mu_o $ and $ \bar{\mu} $ schemes and the natural choice of basis states in each case for deriving the difference equations. To assist comparison with earlier literature, we follow the notation of \cite{Yang2009} in this section. Sec. \ref{von_Neumann_subsection} provides a brief introduction to the method of von Neumann stability of finite difference equations which is used in subsequent sections to understand viability of mLQC-I and mLQC-II. For further details of this method, see for eg. Ref.  \cite{Strikwerda2004}. 

\subsection{Some quantization ambiguities in Hamiltonian constraint} \label{quantizations}

In LQC, in order to pass to the quantum theory, the Hamiltonian formulation of GR is first reformulated in terms of Ashtekar variables -- the connection $ A^i_a $ and the densitized triad $ E_i^a $. Since LQC is the loop quantization of symmetry reduced cosmological models, in case of spatially flat FLRW spacetime, the classical gravitational phase space variables can be symmetry reduced as \cite{Ashtekar2003}
\begin{equation}\label{key}
A_a^i = c V_o^{1/3} \text{ }{^o \omega}^i_a \quad \text{and} \quad E^a_i = p V_o^{-2/3} \sqrt{^o q} \text{ } {^o e}^a_i,
\end{equation}
where $ \lbrace {^oe}^a_i \rbrace$ and $ \lbrace {^o \omega}^i_a \rbrace $ are a set of orthonormal triads and co-triads respectively, and $ V_o $ is the volume of an elementary cell $ {\cal V} $ (with respect to the fiducial flat metric $ ^o q_{ab} $) which is introduced to define a symplectic structure. The Poisson bracket between the symmetry reduced phase space variables is given by,
\begin{equation}\label{key}
\lbrace c,p \rbrace = \frac{\kappa \gamma}{3},
\end{equation} 
where $ \kappa = 8\pi G $ and $ \gamma $ is the Barbero-Immirzi parameter. The classical gravitational Hamiltonian constraint for lapse $ N=1 $ can be written as,
\begin{equation}\label{key}
H_{\mathrm{grav}} = \int_{V} \mathrm{d}^3 x \frac{E^{aj} E^{bk}}{2\kappa \sqrt{\mathrm{det}(q)}} [\epsilon_{ijk}F^i_{ab} - 2(1+\gamma^2)K^j_{\lbrack a} K^k_{b\rbrack}] \equiv H^E - 2(1+\gamma^2)T.
\end{equation}
where we denote by $ H^E $ and $ T $ the Euclidean and the Lorentzian term respectively. In case of spatially flat FLRW spacetime, their expressions reduce to,
\ba
H^E &=& \int_{V} \mathrm{d}^3 x \frac{E^{aj} E^{bk}}{2\kappa \sqrt{\mathrm{det}(q)}}  \epsilon_{ijk}F^i_{ab} = \frac{3}{\kappa} c^2 \sqrt{|p|}, \label{HE} \\
T &=& \int_{V} \mathrm{d}^3 x \frac{E^{aj} E^{bk}}{2\kappa \sqrt{\mathrm{det}(q)}} K^j_{\lbrack a} K^k_{b\rbrack} = \frac{3}{2\kappa \gamma^2} c^2 \sqrt{|p|}. \label{T}
\ea

We note that above expressions are proportional to each other, which allows one to combine them into a single term at the classical level to get $ H_{\mathrm{grav}} = -3 c^2 \sqrt{|p|}/\kappa \gamma^2 $. The loop quantization of this symmetry reduced Hamiltonian constraint where the Lorentzian term is combined with the Euclidean term leads to the standard LQC, with a quantum Hamiltonian constraint as a second order finite difference equation \cite{as-status}. The modified versions of LQC arise when these terms are quantized independently. The quantization obtained from taking the expression \eqref{T} as the starting point is known as the mLQC-I quantization.

In addition, in the spatially flat model one can use 
\begin{equation}\label{KA}
K^i_a = \frac{1}{\gamma} A^i_a .
\end{equation}
to write an alternative expression for the Lorentzian term as:
\begin{equation}\label{alternativeT}
\mathring{T} = \int_{V} \mathrm{d}^3 x \frac{E^{aj} E^{bk}}{2\kappa \sqrt{\mathrm{det}(q)}} A^j_{\lbrack a} A^k_{b\rbrack}. 
\end{equation}
Classically the two expressions \eqref{T} and \eqref{alternativeT} for the Lorentzian terms are equivalent to each other. However, these two expressions for the Lorentzian term  leads to qualitatively different models. The quantization based on the expression \eqref{alternativeT} is called mLQC-II quantization. As mentioned earlier, this quantization only arises for spatially flat spacetimes in contrast to the mLQC-I quantization which arises in general.

To obtain the quantum Hamiltonian constraint, the next step is to express the Hamiltonian constraint in terms of holonomies. For this, let us first note that  the terms involving triads in (\ref{key}) can be rewritten as:
\begin{equation}\label{Thiemann1}
\epsilon_{ijk}\frac{E^{aj} E^{bk}}{\sqrt{\mathrm{det}(q)}} = \frac{2}{\kappa \gamma} \epsilon^{abc} \lbrace A^i_c ,V \rbrace .
\end{equation}
Next, the field strength $ F^i_{ab} $ is represented using the holonomy $ h^{(\lambda)}_{\Box_{ij}} $ around a square loop $ \Box_{ij} $ in the $ i-j $ plane parallel to one of the faces of the elementary cell. The sides of the square are of length $ \lambda V_o^{1/3} $ each with respect to the fiducial metric $ {^o q}_{ab} $. The field strength can then be written as \cite{aps3}:
\begin{equation}\label{fieldstrength}
	F^k_{ab} = -2 \lim_{Ar_{\Box}\to 0} \mathrm{Tr} \bigg(\frac{h^{(\lambda)}_{\Box_{ij}}-1}{\lambda^2 V_o^{2/3}}\bigg) \tau^k \text{ } {^o \omega}^i_a {^o \omega}^j_b 
\end{equation} 
where $ Ar_{\Box} $ is the area of the square loop and $ \tau_i=- i\sigma_i /2 $ where $ \sigma_i $ are Pauli spin matrices. The holonomy $ h^{(\lambda)}_{\Box_{ij}} $ around the square is given by the product of the holonomies along its four edges:
\begin{equation}\label{key}
	h^{(\lambda)}_{\Box_{ij}} = h^{(\lambda)}_i h^{(\lambda)}_j (h^{(\lambda)}_i)^{-1} (h^{(\lambda)}_j)^{-1}
\end{equation}
where the holonomies $ h^{(\lambda)}_i $ of the connection are given by,
\begin{equation}\label{key}
	h_i^{(\lambda)} = \cos\bigg(\frac{\lambda c}{2}\bigg) \mathbb{I} + 2 \sin\bigg(\frac{\lambda c}{2}\bigg) \tau_i, 
\end{equation}
along edges of length $ \lambda $ parallel to the triads $ {^o e}^a_i $. 


Note that the extrinsic curvature can be rewritten using the following identities:
\ba
K^i_a &=& \frac{1}{\kappa \gamma} \lbrace A^i_a,K \rbrace , \label{k-holonomy1} \\
K &=& \gamma^{-2} \lbrace H^E,V \rbrace\label{k-holonomy2}
\ea
and  the Poisson brackets of $ A^i_a $ with $ V $ or $ K $ can be computed using:
\ba
\lbrace c\tau_i , V \rbrace &=& - \frac{1}{\lambda} h^{(\lambda)}_i \lbrace (h^{(\lambda)}_i)^{-1}, V \rbrace, \\
\lbrace c\tau_i , K \rbrace &=& - \frac{1}{\lambda} h^{(\lambda)}_i \lbrace (h^{(\lambda)}_i)^{-1}, K \rbrace . \label{K_holonomy}
\ea
The above expressions suffice if $\lambda$ is a constant such as in the $\mu_o$ quantization of LQC \cite{aps2}, but not if it is a function of triads as in the $\bar \mu$ quantization \cite{aps3}.  In the latter case, $\bar{\mu}^2 |p| = \Delta$, with $ \Delta =4\sqrt{3}\pi \gamma \lp^2$ as the minimum area of the loop in LQG. As a result, the expression \eqref{K_holonomy} needs to take in to account triad dependence of $\lambda$ and one uses:
\begin{equation}\label{K-holonomy2}
\lbrace c\tau_i , K \rbrace = - \frac{2}{3\bar{\mu}} h^{(\bar{\mu})}_i \lbrace (h^{(\bar{\mu})}_i)^{-1}, K \rbrace.
\end{equation}

Using above expressions, it is possible to cast the Hamiltonian constraint completely in terms of the holonomies and the triads or volume $ V = |p|^{3/2} $. In the $ \mu_o $ scheme, the fundamental operators are the triads $ \hat{p} $ and elements of holonomies $ \widehat{\exp(i{\mu_o}c/2)} $, which have a simple action on the eigenstates of the triad operator \cite{Ashtekar2003}:
\be\label{triadev}
\hat{p} |n \rangle = \bigg(\frac{8\pi \gamma \lp^2}{6} \bigg) n |n\rangle, ~~~~
\widehat{\exp(i{\mu_o}c/2)} |n\rangle = |n+\mu_o \rangle . 
\ee
In the $\bar \mu$ quantization, the operator $ \widehat{\exp(i\bar{\mu}c/2)} $ has a complicated action on the basis states $ |p\rangle $ due to dependence of $ \bar{\mu} $ on the triad itself. Interestingly, the action is simple in the volume representation in which the action of $ \widehat{\exp(i\bar{\mu}c/2)} $ yields a uniform difference equation. In particular, in the volume representation,
\be
\hat{V} |v \rangle = \bigg(\frac{8\pi \gamma \lp^2}{6} \bigg)^{3/2} \frac{|v|}{\alpha} |v\rangle, ~~~~
\widehat{\exp(i\bar{\mu}c/2)} |v\rangle = |v+1 \rangle 
\ee
where $\alpha = 2/3\sqrt{3\sqrt{3}}$. 
Using the action of these basic operators, we can calculate the action of the Hamiltonian constraint on the basis states which results in a finite difference equation with discreteness determined by the area gap $\Delta$. As we will discuss in Secs. III and IV the difference equations turn out to be fourth order for mLQC-I and mLQC-II. As in LQC, for $\mu_o$ quantizations, the difference equations are uniform spacing in triad representation, whereas in the $\bar \mu$ quantization they are of uniform discreteness in volume representation.

\subsection{Von Neumann stability analysis of finite difference equations} \label{von_Neumann_subsection}
 In its conventional applications, stability analysis plays an important part in obtaining reliable numerical simulations of PDEs using finite difference methods. The defining feature required of a faithful finite difference scheme is \textit{convergence}, which implies that its solutions should approximate the solutions of the corresponding PDE and that the approximation improves as the step sizes tend to zero. The first step in obtaining a convergent numerical scheme is to  find a \textit{consistent} scheme which ensures that the difference equations approximate the corresponding differential equations as the step sizes decrease. However consistency by itself is not sufficient to ensure convergence. A fundamental theorem in numerical analysis due to Lax and Richtmyer \cite{Lax1956} states that a consistent finite difference scheme for a PDE with a well-posed initial value problem is convergent if and only if it is \textit{stable}. Stability of a finite difference scheme ensures that the numerical errors (with respect to the solutions of the underlying PDE) made at any one time step do not grow exponentially at further steps in the computation. Due to stability, the solutions remain close to the solutions of the PDE under evolution, and approach them in the limit of vanishing discretization. The conditions for stability are in fact closely related to the well-posedness of initial value problem for PDEs \cite{Strikwerda2004}. The von Neumann stability analysis involves Fourier transforming the difference equations. As a result, the stability condition becomes a condition on the roots of an algebraic equation obtained from Fourier transform of the difference equation. As an illustration, consider the simplest case of a linear PDE in two variables with constant coefficients:
\begin{equation}\label{key}
\frac{\partial f(x,t)}{\partial t} = a \frac{\partial f(x,t)}{\partial x}.
\end{equation}

This can be discretized using a one-step finite difference scheme as follows:
\begin{equation}\label{key}
\frac{1}{h_t}\bigg(|m,n+1\rangle - |m,n\rangle\bigg) = \frac{a}{h_x}\bigg(|m,n\rangle-|m-1,n\rangle\bigg),
\end{equation}
where $ h_t $ and $ h_x $ are constants that represent the discretization in time and space respectively, and $ |m,n\rangle $ represents the value of the solution on a discrete set of points where $ n $ labels the discrete time slices and $ m $ represents discrete spatial points in each time slice. The above difference equation can be recast to express the value of the function at time slice $ n+1 $ in terms of its values at various spatial points at time slice $ n $:
\begin{equation}\label{key}
|m,n+1\rangle = (1+a')|m,n\rangle - a'|m-1,n\rangle,
\end{equation}
where $ a'=ah_t /h_x $ is a constant that depends on the discretizations in both $ t $ and $ x $. Stability requires that the norm of the solution at any time $ t $ is limited in the amount of growth from its initial value. The von Neumann analysis allows us to express this condition in terms of a simple algebraic condition involving the discretization parameters by performing a Fourier transform of the difference equation. Performing the Fourier transform in the spatial coordinate, reduces the difference equation to,
\begin{equation}\label{amplieq}
f^{n+1} (s) = \bigg[(1-a')+a' e^{-i h_x s }\bigg] f^n (s) = g(h_x s)f^n (s),
\end{equation}
where the amplification factor $ g $ relates the value of the solution at one time step to its value at the next time step. It is important to note that all solutions of a finite difference equation of the above form satisfy $f^n = g^n f^0$. Further, to determine the amplification factor $g$ an equivalent procedure to Fourier analysis is to substitute $|m,n\rangle = g^n e^{ims} $ in the difference equation which quickly yields an algebraic equation for the amplification factor $ g $. As rigorously shown in Ref. \cite{Strikwerda2004}, using this ansatz must not be seen as seeking solutions of the difference equations and understanding its properties for this particular ansatz, but instead the procedure is completely equivalent to the method based on Fourier transforms without taking any ansatz. This shortcut allows one to understand stability properties of the difference equation in an easy and straightforward way. The von Neumann stability condition requires that we must have $ |g| \leq 1 $ to suitably limit the growth of the numerical solution \cite{Strikwerda2004}. In case of higher oder PDEs, we get a polynomial equation in the amplification factor $ g $ with multiple roots. Stability conditions then require that $ |g| \leq 1 $ for all the roots of the polynomial equation.\footnote{There can be situations in which physical solutions are obtained only from a set of roots, while other roots do not contribute. In such a case, absolute value of amplification factor for the physical relevant roots must be bounded by unity. A situation such as this arises in Sec. IV.}  

However, as mentioned earlier, the application of von Neumann stability analysis to difference equations of loop quantum cosmologies differs from its conventional applications in motivation. In the conventional applications, the differential equations are fundamental and the difference equations are used to approximate the former. The situation is opposite in loop quantum cosmological models. Here the difference equations arising due to quantum disceretness are fundamental, but they are well approximated 
by the differential Wheeler-DeWitt equation in the large volume limit where the quantum discreteness scale can be ignored \cite{Bojowald2002,Ashtekar2003,ps12}. This is straightforward to show by analyzing the quantum Hamiltonian constraint in loop quantum cosmological models at large volumes for semi-classical states. In Appendix A, we summarize this analysis for the case of one of the loop quantum Hamiltonian constraints considered in our analysis. Further, solutions of LQC approximate those of the Wheeler-DeWitt equation in the large volume limit \cite{Ashtekar2003,aps2,aps3}.   Since the semi-classical solutions of the Wheeler-DeWitt theory approximate GR excellently in the infrared limit, if the semi-classical solutions of the loop quantum difference equations approximate closely the solution of the Wheeler-DeWitt differential equations we are ensured that the correct classical limit is obtained at large volumes. We apply von Neumann stability analysis to the difference equations in loop quantized spacetimes in this context to understand whether at large volumes one obtains a good approximation to the Wheeler-DeWitt equation.

Consequently, there are some technical differences in the application of the method and the interpretation of results in loop quantum cosmologies. In the standard application of von Neumann analysis, the difference equation is expected to approximate the differential equation in the limit of vanishing discretization parameter. However in loop quantum cosmologies, the discreteness in variables emerges from the underlying discreteness of the quantum geometry. The underlying Planck scale discreteness of the quantum spacetime is fixed and there is no freedom to change it to improve numerical efficiency or to take the limit of vanishing discretization. In fact, we expect the difference equation to depart radically from the differential equation in the Planck regime due to quantum gravity effects. We thus use the von Neumann analysis in loop quantum cosmologies only in large volume limit where the discreteness scale can effectively be ignored. This is also the limit in which the difference equation in loop quantum cosmologies is expected to approximate the Wheeler-DeWitt equation (see Appendix A). Further, we are interested in analyzing the behavior of the von Neumann amplification factor with respect to volume, i.e. the spatial coordinates, thus the Fourier transform is performed in $ \phi $, the massless scalar field which acts as the internal time coordinate. This is in contrast to the standard application of the method where the Fourier transform is performed in spatial coordinates and the amplification factor relates solutions from one time step to the next. Also note that the von Neumann analysis is conventionally only meant for difference equations with constant coefficients. One can apply this method to difference equations with variable coefficients, as is usually the case in loop quantum cosmologies, using the frozen coefficient approximation. In this approximation, stability is considered in a small neighborhood where the coefficients in the PDE can be considered to be constant.

Von Neumann analysis has been studied in standard LQC for various settings (for a review see Ref. \cite{khanna-review,ps12}).  It was first used in Refs. \cite{Bojowald2003,date2}, and has been applied to study difference equations in anisotropic models \cite{Cartin2005,Cartin2005a,Rosen2006,martin-cartin-khanna,Nelson2009} and black hole spacetimes \cite{Cartin2006,Yonika2018}. It is interesting to note that several phenomenological problems with some of the LQC quantizations such as $\mu_o$ prescription in presence of matter violating strong energy condition \cite{cs-unique}, can be independently deduced from the stability analysis. The instability of quantum difference equation reveals many details which confirm phenomenological problems as well as constraints on the solutions in the physical Hilbert space \cite{ps12}. In the following, we will use these techniques to understand properties of finite difference equations arising in modified loop quantum cosmologies.

\section{von Neumann stability analysis of \lowercase{m}LQC-I quantization} \label{mLQC-I}

In mLQC-I one quantizes the expression \eqref{T} for the Lorentzian term using the relations \eqref{k-holonomy1} and \eqref{k-holonomy2} to express the extrinsic curvature in terms of the holonomies. This results in a fourth order difference equation whose von Neumann stability for $\mu_o$ and $\bar \mu$ prescriptions will be studied in the following. We will consider cases with massless scalar field, and of massless scalar field with a cosmological constant. These two cases span the entire range of the weak energy condition for matter content, and in LQC provided important insights on the viability of $\bar \mu$ quantization and the limitations of the $\mu_o$ approach. We will find in Sec. \ref{mLQC-I-1} that the difference equation turns  out to be stable in the $ \mu_o $ scheme if we are only considering free scalar field as our matter Hamiltonian. However, when we add a positive cosmological constant to the matter Hamiltonian (discussed in Sec. \ref{mLQC-I-2}), it becomes unstable. This shows the limitation of the $ \mu_o $ scheme in mLQC-I quantization. We will find in Sec. \ref{mLQC-I-3} and \ref{mLQC-I-4} that the mLQC-I in the $ \bar{\mu} $ scheme is von Neumann stable for both the scalar field case as well as the case when a  positive cosmological constant is included. This provides a strong hint for the viability of this scheme in general.

\subsection{mLQC-I with scalar field in the $ \mu_o $ scheme} \label{mLQC-I-1}

The difference equation for  this quantization of the FLRW model in the $ \mu_o $ scheme was first derived in Ref. \cite{Bojowald2002} in the triad basis, where it was analyzed for the vacuum case. Here we outline the derivation of the difference equation below. In the interest of maintaining uniformity of notation in this manuscript, we will follow the notation used in \cite{Yang2009}.

In order to obtain the operator for the Euclidean term, we first write the expression \eqref{HE} in terms of holonomies using the expressions \eqref{Thiemann1} and \eqref{fieldstrength}. We elevate the volume and the holonomies to operators and obtain the expression for the Hamiltonian constraint operator in the $ \mu_o $ scheme. The operator form of the Euclidean part can be shown to be:
\begin{equation}\label{euclidean_mu_o}
H^{E,\mu_o} = - \frac{24 \gamma^2  i}{2 \kappa^2 \hbar \gamma^3} \sin(\mu_o c) \frac{\mathrm{sgn}(v)}{\mu_o^3}\bigg(\sin\bigg(\frac{\mu_o c}{2}\bigg) V \cos\bigg(\frac{\mu_o c}{2}\bigg) - \cos\bigg(\frac{\mu_o c}{2}\bigg) V \sin\bigg(\frac{\mu_o c}{2}\bigg)\bigg) \sin(\mu_o c),
\end{equation}
where we have omitted the hats on the operators for brevity. We require eqs. \eqref{k-holonomy1} and \eqref{k-holonomy2} in addition to \eqref{Thiemann1} to express the Lorentzian term \eqref{T} in terms of holonomies, which yields
\ba
T^{\mu_o} &=& - \frac{24 i}{\kappa^4 \hbar^5 \gamma^7} \bigg(\sin\bigg(\frac{\mu_o c}{2}\bigg) B \cos\bigg(\frac{\mu_o c}{2}\bigg) - \cos\bigg(\frac{\mu_o c}{2}\bigg) B \sin\bigg(\frac{\mu_o c}{2}\bigg)\bigg) \nonumber \\
&& \times \frac{\mathrm{sgn}(v)}{\mu_o^3}\bigg(\sin\bigg(\frac{\mu_o c}{2}\bigg) V \cos\bigg(\frac{\mu_o c}{2}\bigg) - \cos\bigg(\frac{\mu_o c}{2}\bigg) V \sin\bigg(\frac{\mu_o c}{2}\bigg)\bigg) \nonumber \\
&& \times \bigg(\sin\bigg(\frac{\mu_o c}{2}\bigg) B \cos\bigg(\frac{\mu_o c}{2}\bigg) - \cos\bigg(\frac{\mu_o c}{2}\bigg) B \sin\bigg(\frac{\mu_o c}{2}\bigg)\bigg), \label{Lorentzian_mu_o}
\ea
where we have defined the operator $ B = [H^{E,\mu_o}, V] $. The resulting action of the Hamiltonian constraint is given by
\begin{equation}\label{key}
H_{\mathrm{grav}} |n\rangle + H_{\phi} |n\rangle = (H^{E,\mu_o} - 2(1+\gamma^2)T^{\mu_o}) |n\rangle + H_{\phi} |n\rangle = 0 .
\end{equation}
Using the operators for $ H^{E,\mu_o} $ and $ T^{\mu_o} $ as mentioned above, we obtain the following difference equation,
\ba
\frac{9}{2}(1+\gamma^2) Q_1 F_1(n) |n+8\mu_o \rangle - \frac{\gamma^2 Q_2}{2} W(3,1) |n+4\mu_o \rangle && \nonumber \\
+ \left\{ \frac{9}{2}(1+\gamma^2) Q_1 F_2(n) - \frac{\gamma^2 Q_2}{2} (W(1,3)+W(-3,-1)) - \frac{B(n)}{2} \partial_{\phi}^2 \right\} |n\rangle && \nonumber \\
- \frac{\gamma^2 Q_2}{2} W(-1,-3) |n-4\mu_o \rangle + \frac{9}{2}(1+\gamma^2) Q_1 F_3(n) |n-8\mu_o \rangle &=& 0
\ea
where we have used the following abbreviated notation,
\ba
W(a,b) &=& |n+a\mu_o|^{3/2} - |n+b\mu_o|^{3/2}, \\
F_1(n) &=& [W(5,9)W(8,6)-W(3,7)W(6,4)] W(5,3) \nonumber \\
&& \times [W(1,5)W(4,2)-W(-1,3)W(2,0)], \\
F_2(n) &=& [W(5,1)W(4,2)-W(3,-1)W(2,0)] W(5,3) \nonumber \\
&& \times [W(1,5)W(4,2)-W(-1,3)W(2,0)] \nonumber \\
&& + [W(-3,1)W(0,-2)-W(-5,-1)W(-2,-4)] W(-3,-5) \nonumber \\
&& \times [W(1,-3)W(0,-2)-W(-1,-5)W(-2,-4)], \\
F_3(n) &=& [W(-3,-7)W(-4,-6)-W(-5,-9)W(-6,-8)] W(-3,-5) \nonumber \\
&& \times [W(1,-3)W(0,-2)-W(-1,-5)W(-2,-4)],
\ea 
and the coefficient $ B(n) $  given by \cite{aps1},
\begin{equation}\label{key}
B(n) = \bigg(\frac{8\pi \gamma \lp^2}{6} \bigg)^{-3/2} \bigg(\frac{2}{3\mu_o}\bigg)^6 \bigg[|n+\mu_o|^{3/4} - |n-\mu_o|^{3/4}\bigg]^6.
\end{equation}
Above, $ Q_1 $ and $ Q_2 $ are following constants:
\ba
Q_1 &=& \frac{1}{\kappa^8 \gamma^9 \hbar^7 \mu_o^9} \bigg(\frac{8\pi \gamma \lp^2}{6} \bigg)^{15/2}, \\
Q_2 &=& \frac{3}{\kappa^2 \gamma^3 \hbar \mu_o^3} \bigg(\frac{8\pi \gamma \lp^2}{6} \bigg)^{3/2}.
\ea

We are now set to perform the von Neumann stability analysis. For this we use the ansatz of the form discussed below eq.(\ref{amplieq}):
\begin{equation}\label{ansatz}
|n\rangle = g^{n/4\mu_o} e^{i\omega \phi}
\end{equation}
As noted in previous section, using this particular ansatz in completely equivalent to performing von-Neumann stability analysis using Fourier integrals and is only a short-cut to obtain results on stability \cite{Strikwerda2004}. Hence, any results following from using this ansatz (or a similar ansatz in the following sections) should not be seen tied to the choice of the ansatz. 
Using this ansatz in the above difference equation, we obtain a polynomial equation for the amplification factor,
\ba
\frac{9}{2}(1+\gamma^2) Q_1 F_1(n) g^2 - \frac{\gamma^2 Q_2}{2} W(3,1) g && \nonumber \\
+ \left\{ \frac{9}{2}(1+\gamma^2) Q_1 F_2(n) - \frac{\gamma^2 Q_2}{2} (W(1,3)+W(-3,-1)) + \frac{B(n)\omega^2}{2} \right\} && \nonumber \\
- \frac{\gamma^2 Q_2}{2} W(-1,-3) g^{-1} + \frac{9}{2}(1+\gamma^2) Q_1 F_3(n) g^{-2} &=& 0
\ea
In the large volume limit, the above polynomial equation reduces to,
\begin{equation}\label{rootsTmuo}
R g^4 - g^3 + 2(1-R) g^2 - g + R = 0,
\end{equation}
where $ R := (1+ \gamma^2)/4\gamma^2 $. The above equation has four roots which come in pairs. Two of them are real and both equal to unity. And the other two are a complex pair of roots $ (\gamma - i)/(\gamma+i), (\gamma+i)/(\gamma-i) $ which are complex conjugate of each other. All the roots are of unit magnitude, so the difference equation in this case is von Neumann stable in the large volume limit. In contrast, difference equation in $\mu_o$ quantization for massless scalar field in LQC results in two roots of unit magnitude. Though unit magnitude of roots signals viability of quantization, we should be careful in reaching any conclusion as it is important to test the viability by inclusion of other matter such as cosmological constant which we consider in the following.

\subsection{mLQC-I with scalar field and a positive cosmological constant in the $ \mu_o $ scheme} \label{mLQC-I-2}

\begin{figure}
	\includegraphics[width=.4\textwidth]{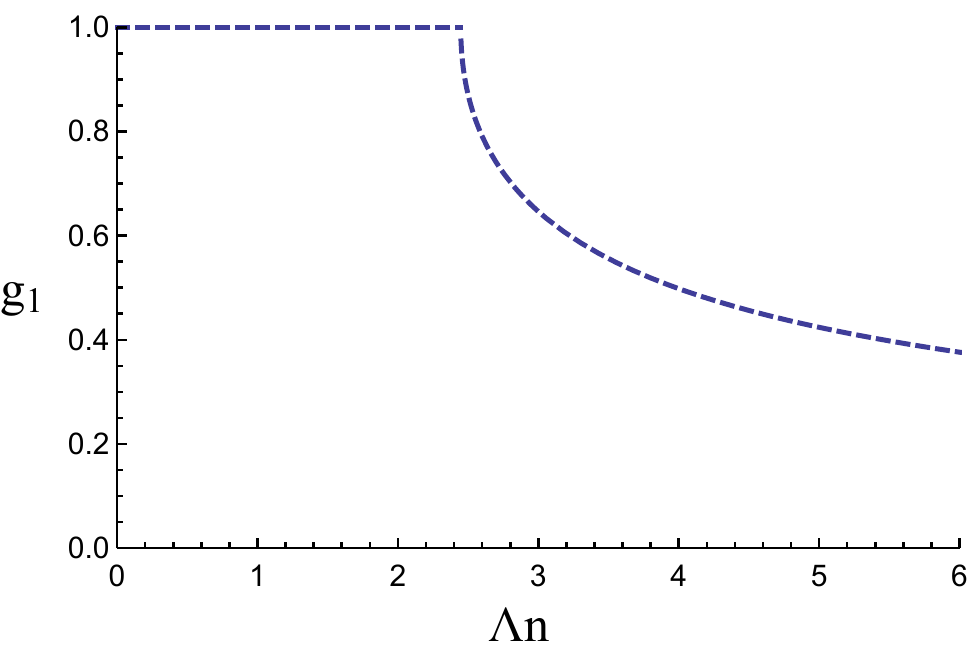}
	\includegraphics[width=.4\textwidth]{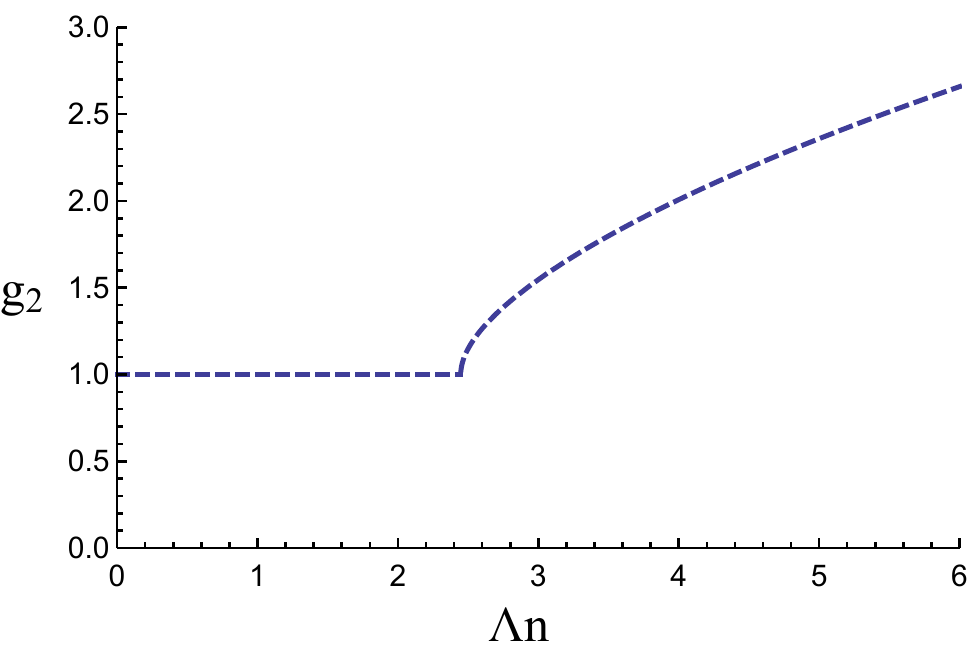}
	\includegraphics[width=.4\textwidth]{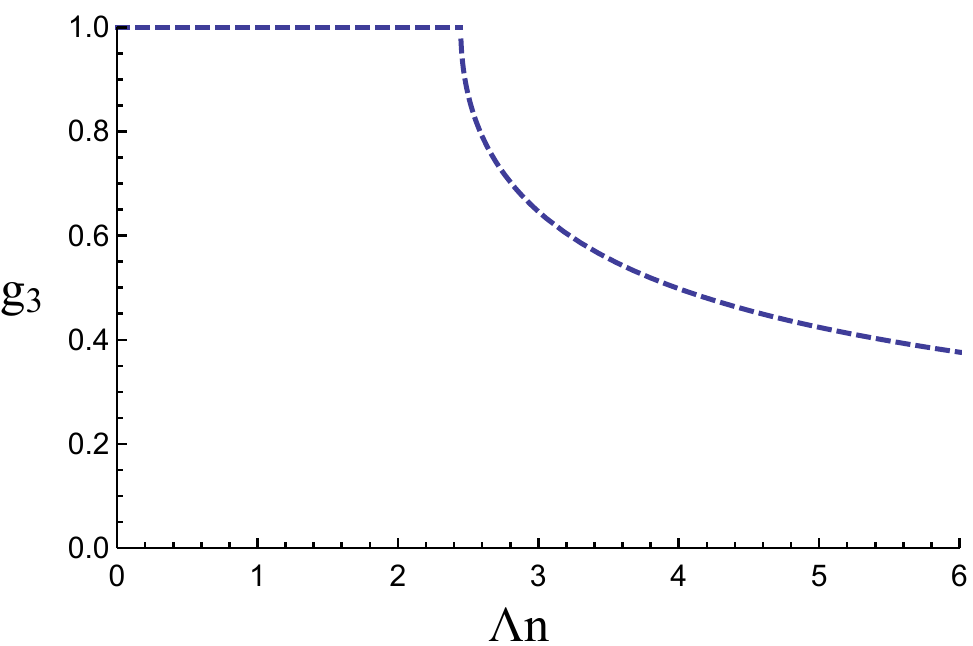}
	\includegraphics[width=.4\textwidth]{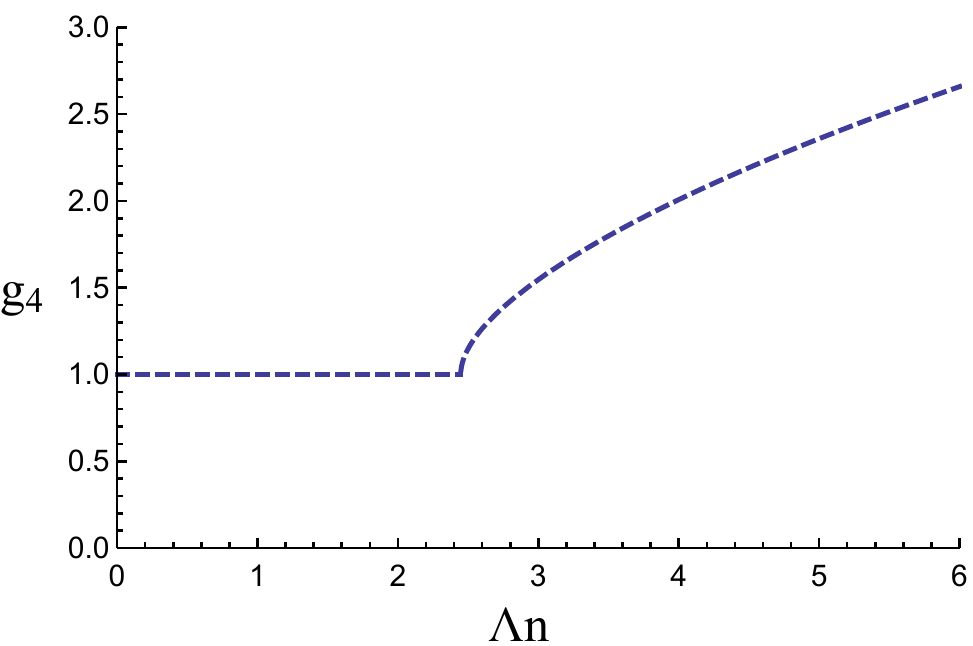}
	\caption{Absolute values of the roots are plotted for mLQC-I with scalar field and a positive cosmological constant in the $ \mu_o $ scheme.}
	\label{fig:fig1}
\end{figure}

To test the viability of any scheme we must explore different choices of matter content. The main drawback of the $\mu_o$ scheme in LQC resulted from including matter which violates strong energy condition \cite{cs-unique}, which is best understood by inclusion of a positive cosmological constant. To understand the viability of $\mu_o$ scheme in mLQC-I, we consider a matter Hamiltonian with a positive cosmological constant,
\begin{equation}\label{Hmatter_cosmologicalconstant}
H_{\mathrm{m}} = H_{\phi} + \frac{\Lambda}{\kappa}V.
\end{equation}
The difference equation resulting from quantum Hamiltonian constraint then becomes,
\ba
\frac{9}{2}(1+\gamma^2) Q_1 F_1(n) |n+8\mu_o \rangle - \frac{\gamma^2 Q_2}{2} W(3,1) |n+4\mu_o \rangle && \nonumber \\
+ \left\{ \frac{9}{2}(1+\gamma^2) Q_1 F_2(n) - \frac{\gamma^2 Q_2}{2} (W(1,3)+W(-3,-1)) \right. && \nonumber \\
\left. - \frac{B(n)}{2} \partial_{\phi}^2 \right\} |n\rangle - \frac{\gamma^2 Q_2}{2} W(-1,-3) |n-4\mu_o \rangle && \nonumber \\
+ \frac{9}{2}(1+\gamma^2) Q_1 F_3(n) |n-8\mu_o \rangle &=& -\bigg(\frac{8\pi \gamma \lp^2}{6} \bigg)^{3/2} \frac{\Lambda}{\kappa} |n|^{3/2} |n\rangle.
\ea

Using the ansatz (\ref{ansatz}), this difference equation at large volume reduces to the following polynomial equation for the amplification factor:
\begin{equation}
R g^4 - g^3 + 2\bigg(1-R + \frac{\mu_o^2 \kappa \hbar \gamma}{9} \Lambda n\bigg) g^2 - g + R = 0,
\end{equation}
The roots of this equation in Planck units become,
\ba
g_{1,2} &=& 1-\frac{1}{1+\gamma^2}\bigg[1-\frac{1}{3}\sqrt{9-2R_1 \gamma^2 \Lambda n} \pm \frac{\sqrt{2}\gamma}{3}\sqrt{-9-R_1 \Lambda n + 3 \sqrt{9-2R_1 \gamma^2 \Lambda n}} \bigg], \\
g_{3,4} &=& 1-\frac{1}{1+\gamma^2}\bigg[1+\frac{1}{3}\sqrt{9-2R_1 \gamma^2 \Lambda n} \pm \frac{\sqrt{2}\gamma}{3}\sqrt{-9-R_1 \Lambda n - 3 \sqrt{9-2R_1 \gamma^2 \Lambda n}} \bigg],
\ea
where $ R_1 := 32\sqrt{3}\pi^2 \gamma^2 (1+\gamma^2) $. In the limit of vanishing $ \Lambda $, we get back the roots of equation \eqref{rootsTmuo}. The pair $ g_1, g_2 $ corresponds to the pair of real roots in equation \eqref{rootsTmuo}, and the pair $ g_3, g_4 $ corresponds to the complex pair of roots. Note that the roots depend on the combination $ \Lambda n $ and not just $ \Lambda $. In order to analyze the stability of the underlying difference equation, we consider the behavior of the absolute value of the roots as a function of $ \Lambda n $,  shown in Fig. \ref{fig:fig1}. We see that one of the roots from each pair, namely $ g_2 $ and $ g_4 $ have magnitudes larger than unity for values of $ \Lambda n $ exceeding a certain finite value. For any given value of $ \Lambda $, the combination $ \Lambda n $ can be made as large as desired by choosing a large enough volume, such that two of the roots have magnitude larger than unity and each of the pair of roots is problematic. Thus we conclude that the difference equation for mLQC-I in the $ \mu_o $ scheme is von Neumann unstable in the presence of a positive cosmological constant. Since none of the pair of roots above results in von Neumann stability, this is a strong evidence that this quantization in $ \mu_o $ scheme is physically non-viable in presence of a positive cosmological constant at large volumes.

\subsection{mLQC-I with scalar field in the $ \bar{\mu} $ scheme} \label{mLQC-I-3}
Problems of $\mu_o$ scheme in LQC, like the one we just encountered above,  were resolved in the $\bar \mu$ quantization. In mLQC-I, 
the difference equation for this case has been obtained in Ref. \cite{Yang2009} for lapse $ N=1 $, and in Ref. \cite{Dapor2017} for lapse $ N=2V $. The von Neumann stability analysis in both these cases leads to the same equation for the amplification factor $ g $ in both scenarios - when only scalar field is present as well as in the presence of  a positive cosmological constant. Here we provide details for only one of them, the one with lapse $ N=1 $ in the notation used in Ref. \cite{Yang2009}. The expression for the Euclidean and Lorentzian part of the Hamiltonian constraint in $ \bar{\mu} $ scheme can be written as follows:
\ba
H^{E,\bar{\mu}} &=& - \frac{24 \gamma^2  i}{2 \kappa^2 \hbar \gamma^3} \sin(\bar{\mu} c) \frac{\mathrm{sgn}(v)}{\bar{\mu}^3}\bigg(\sin\bigg(\frac{\bar{\mu} c}{2}\bigg) V \cos\bigg(\frac{\bar{\mu} c}{2}\bigg) \nonumber \\
&& - \cos\bigg(\frac{\bar{\mu} c}{2}\bigg) V \sin\bigg(\frac{\bar{\mu} c}{2}\bigg)\bigg) \sin(\bar{\mu} c), \label{Euclidean_barmu} \\
T^{\bar{\mu}} &=& - \frac{96 i}{9 \kappa^4 \hbar^5 \gamma^7 } \bigg(\sin\bigg(\frac{\bar{\mu} c}{2}\bigg) B \cos\bigg(\frac{\bar{\mu} c}{2}\bigg) - \cos\bigg(\frac{\bar{\mu} c}{2}\bigg) B \sin\bigg(\frac{\bar{\mu} c}{2}\bigg)\bigg) \nonumber \\
&& \times \frac{\mathrm{sgn}(v)}{\bar{\mu}^3} \bigg(\sin\bigg(\frac{\bar{\mu} c}{2}\bigg) V \cos\bigg(\frac{\bar{\mu} c}{2}\bigg) - \cos\bigg(\frac{\bar{\mu} c}{2}\bigg) V \sin\bigg(\frac{\bar{\mu} c}{2}\bigg)\bigg) \nonumber \\
&& \times \bigg(\sin\bigg(\frac{\bar{\mu} c}{2}\bigg) B \cos\bigg(\frac{\bar{\mu} c}{2}\bigg) - \cos\bigg(\frac{\bar{\mu} c}{2}\bigg) B \sin\bigg(\frac{\bar{\mu} c}{2}\bigg)\bigg), \label{Lorentzian_barmu}
\ea
where we have retained the factor ordering used in Ref. \cite{Yang2009}. Note that $ 1/{\bar{\mu}^3} $ now acts as an operator and is proportional to the volume operator. The pre-factor of constants in the  operator  for the Lorentzian term  is different than in the $ \mu_o $ scheme because of the difference in equations \eqref{K_holonomy} and \eqref{K-holonomy2}. 

As discussed in the previous section, the natural choice of basis states in the $ \bar{\mu} $ scheme is the volume representation. The Hamiltonian constraint in this case leads to the following difference equation \cite{Yang2009},
\begin{eqnarray}
\frac{C_2}{2} |v| \bigg||v+1|^{1/3} - |v-1|^{1/3} \bigg|^3 \partial_{\phi}^2 |v\rangle &=& F'_{+} (v)|v+8\rangle + f'_{+} (v)|v+4\rangle \nonumber\\
&& + (F'_o (v) + f'_o (v))|v\rangle \nonumber\\
&& + f'_{-} (v)|v-4\rangle + F'_{-} (v)|v-8\rangle,  
\end{eqnarray}
where the coefficients are given by,
\begin{eqnarray}
f'_{+} (v) &=& -C_3 (v+2)(|v+3|-|v+1|), \\
f'_{-} (v) &=& f'_{+} (v-4), \\
f'_o (v) &=& -f'_{+} (v)- f'_{-} (v) ,\\
F'_{+} (v) &=& \frac{4 \kappa^2 C_1}{\gamma^4} [M_v (1,5)f'_{+} (v+1)- M_v (-1,3)f'_{+} (v-1)] \nonumber \\
&& \times (v+4) M_v (3,5) [M_v (5,9)f'_{+} (v+5) - M_v (3,7)f'_{+} (v+3)] , \\
F'_{+} (v) &=& \frac{4 \kappa^2 C_1}{\gamma^4} [M_v (1,-3)f'_{-} (v+1)- M_v (-1,-5)f'_{-} (v-1)] \nonumber \\
&& \times (v-4) M_v (-5,-3) [M_v (-3,-7)f'_{-} (v-3) - M_v (-5,-9)f'_{-} (v-5)], \\
F'_o (v) &=& \frac{4 \kappa^2 C_1}{\gamma^4} [M_v (1,5)f'_{+} (v+1)- M_v (-1,3)f'_{+} (v-1)] \nonumber \\
&& \times (v+4) M_v (3,5) [M_v (5,1)f'_{-} (v+5) - M_v (3,-1)f'_{-} (v+3)] \nonumber \\
&& + \frac{4 \kappa^2 C_1}{\gamma^4} [M_v (1,-3)f'_{-} (v+1)- M_v (-1,-5)f'_{-} (v-1)] \nonumber \\
&& \times (v-4) M_v (-5,-3) [M_v (-3,1)f'_{+} (v-3) - M_v (-5,-1)f'_{+} (v-5)].
\end{eqnarray}

Here $ M_v (a,b) $ is defined as
\begin{equation}
M_v (a,b) :=  |v+a| - |v+b|,
\end{equation}
and $ C_1, C_2 $ and $ C_3 $ are constants given by,
\begin{eqnarray}
C_1 &=& 2(1+\gamma^2) \frac{\sqrt{6} \gamma^{\frac{3}{2}}}{2^8 3^3 \kappa^{\frac{3}{2}} \hbar^{\frac{1}{2}} \alpha} ,\\
C_2 &=& \bigg(\frac{3}{2} \bigg)^3 \bigg(\frac{6}{\kappa \hbar \gamma} \bigg)^{3/2} \alpha ,\\
C_3 &=& \frac{\gamma^2}{2 \kappa} \frac{27}{16} \sqrt{\frac{8\pi}{6}} \frac{\alpha \lp}{\gamma^{\frac{3}{2}}}.
\end{eqnarray}

Using the  ansatz:
\begin{equation}\label{ansatz4}
|v\rangle = g^{\frac{v}{4}} e^{i\omega \phi},  
\end{equation}
the von Neumann stability analysis results in the following polynomial equation for the amplification factor $ g $:
\begin{eqnarray}
F'_{+} (v) g^4 + f'_{+} (v) g^3 + \bigg(F'_o (v) + f'_o (v) + C_2 \frac{\omega^2}{2} |v| \bigg||v+1|^{1/3} - |v-1|^{1/3} \bigg|^3  \bigg) g^2 
+f'_{-} (v) g + F'_{-} (v) \nonumber \\ = 0. ~~~~&&~
\end{eqnarray}
In the large volume limit, the above polynomial equation reduces to,
\begin{equation}
R g^4 - g^3 + 2(1-R) g^2 - g + R = 0,
\end{equation}
which is exactly the same as equation \eqref{rootsTmuo}. As discussed earlier, this equation has two pairs of roots, two real and two complex, all having magnitude unity. Thus, the difference equation is von Neumann stable.  We discuss below whether the difference equation in mLQC-I for the $ \bar{\mu} $ scheme is stable even when we add a positive cosmological constant to the matter Hamiltonian.

\subsection{mLQC-I with scalar field and a positive cosmological constant in the $ \bar{\mu} $ scheme} \label{mLQC-I-4}

As noted before in Sec. IIIB, adding a positive cosmological constant to the matter part modifies the matter Hamiltonian as per equation \eqref{Hmatter_cosmologicalconstant}. This will lead to a modification of the difference equation, which in this case becomes,
\begin{eqnarray}
\frac{C_2}{2} |v| \bigg||v+1|^{1/3} - |v-1|^{1/3} \bigg|^3 \partial_{\phi}^2 |v\rangle + \frac{\gamma \lp^2 \sqrt{\Delta}}{4} \Lambda v |v\rangle &=& F'_{+} (v)|v+8\rangle + f'_{+} (v)|v+4\rangle \nonumber\\
&& + (F'_o (v) + f'_o (v))|v\rangle + f'_{-} (v)|v-4\rangle \nonumber\\
&&  + F'_{-} (v)|v-8\rangle.  
\end{eqnarray}

\begin{figure}
	\includegraphics[width=.4\textwidth]{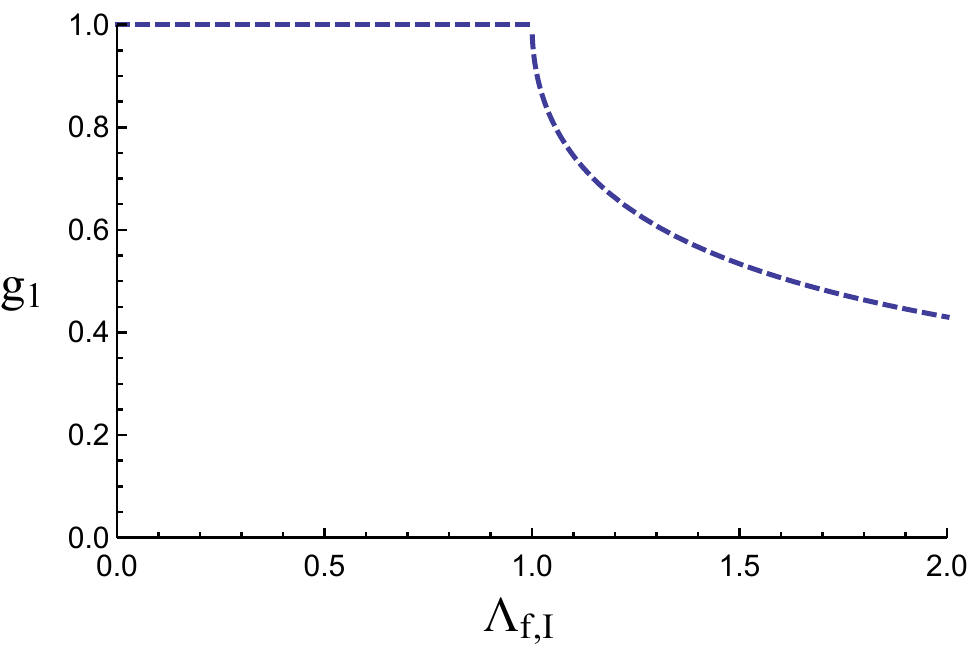}
	\includegraphics[width=.4\textwidth]{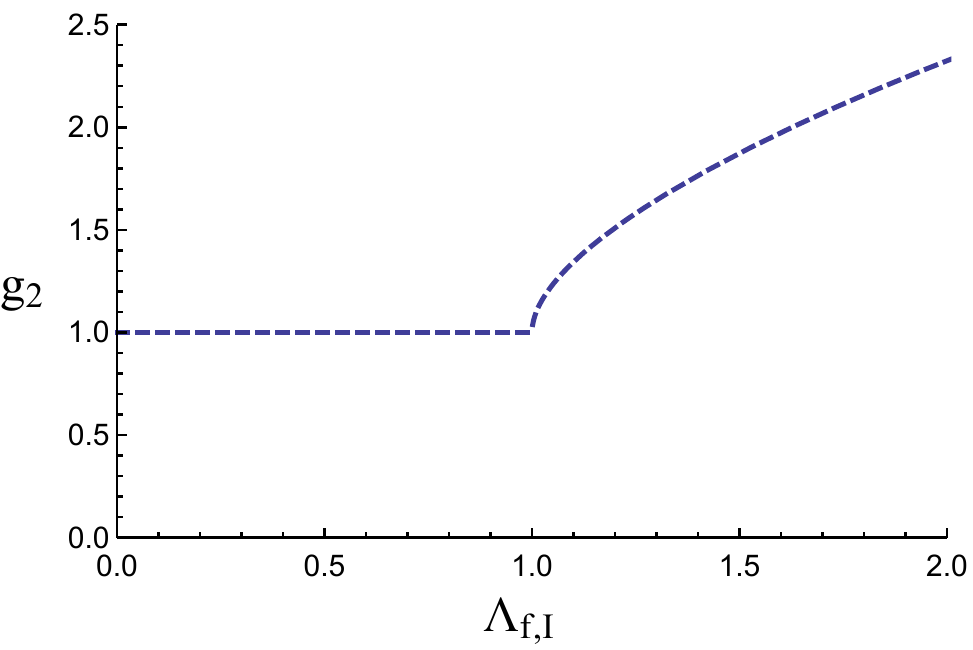}
	\includegraphics[width=.4\textwidth]{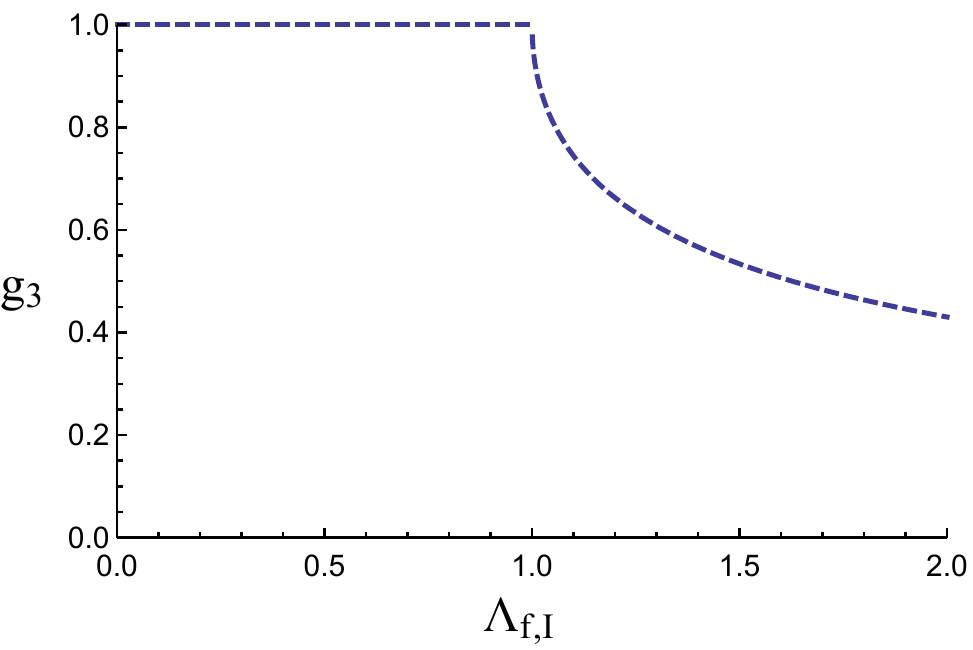}
	\includegraphics[width=.4\textwidth]{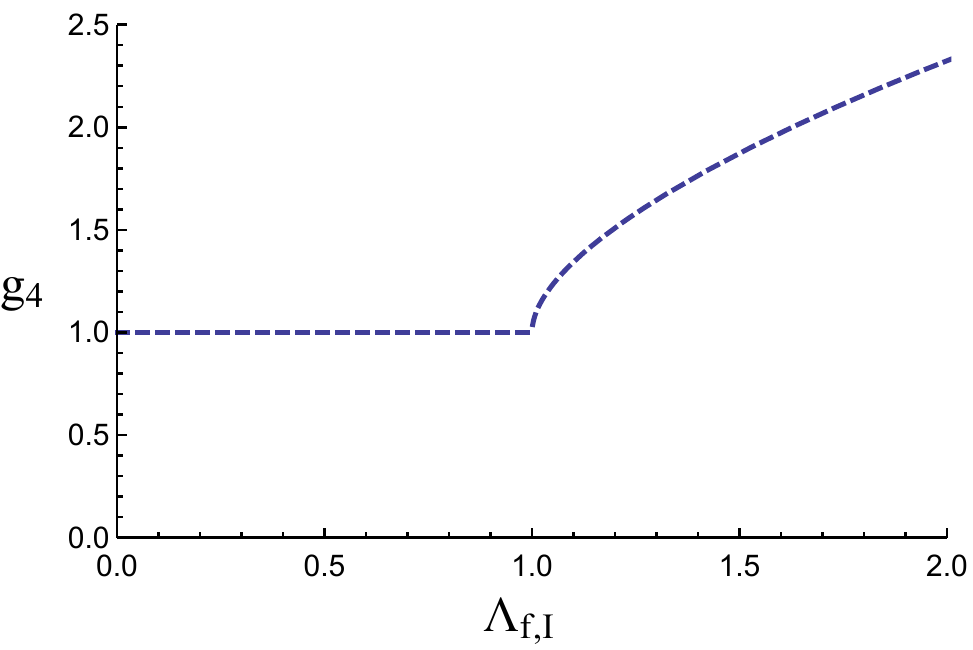}
	\caption{Absolute values of the roots are plotted for mLQC-I with scalar field and a positive cosmological constant in the $ \bar{\mu} $ scheme.}
	\label{fig:fig2}
\end{figure}

Using the ansatz \eqref{ansatz4}, the above difference equation results in the following polynomial equation at large volumes,
\begin{equation}
R g^4 - g^3 + \bigg(2(1-R)+ \frac{1}{4R\gamma^2}\Lambda_{f,\mathrm{I}} \bigg) g^2 - g + R = 0,
\end{equation}
where $\Lambda_{f,\mathrm{I}} := \Lambda/\Lambda_{c,\mathrm{I}} $, and $\Lambda_{c,\mathrm{I}}$ is the critical value of the cosmological constant, which is $ \kappa $ times the critical energy density in mLQC-I. The critical density is the maximum value of energy density  in effective dynamics given by $ 3/(32(\gamma^2 +1)\pi G \gamma^2 \Delta) $ \cite{Yang2009,Li2018}.
The above equation has two pairs of roots,
\ba 
g_{1,2} = \frac{1}{1+\gamma^2}\bigg[\gamma^2 + \sqrt{1-\Lambda_{f,\mathrm{I}}} \mp \sqrt{-2\gamma^2 - \Lambda_{f,\mathrm{I}} +2\gamma^2 \sqrt{1-\Lambda_{f,\mathrm{I}}}} \bigg], \\
g_{3,4} = \frac{1}{1+\gamma^2}\bigg[\gamma^2 - \sqrt{1-\Lambda_{f,\mathrm{I}}} \mp \sqrt{-2\gamma^2 - \Lambda_{f,\mathrm{I}} -2\gamma^2 \sqrt{1-\Lambda_{f,\mathrm{I}}}} \bigg] .
\ea
 The pair $ (g_1,g_2) $ and $ (g_3,g_4) $ reduce to the real and complex pair of roots of equation \eqref{rootsTmuo} respectively when $ \Lambda $ vanishes. As per the analysis of effective dynamics of mLQC-I in $ \bar{\mu} $ scheme Ref. \cite{Li2018}, the Friedmann equation in this case implies that the energy density cannot exceed the critical value of energy density, which in turn implies that $ \Lambda_{f,\mathrm{I}} $ can only vary from $ 0 $ to $ 1 $. Thus, the difference equation is expected to become unstable for $ \Lambda_{f,\mathrm{I}} > 1 $. As shown in Fig. \ref{fig:fig2}, the magnitude of the roots remains unity in the whole range $ 0 \leq \Lambda \leq \Lambda_{c,\mathrm{I}} $, and two of the roots, one from each pair, indeed become greater than unity for $ \Lambda \geq \Lambda_{c,\mathrm{I}} $.

We conclude from this section that the difference equations of mLQC-I, which were found to be unstable in the $ \mu_o $ scheme for scalar field with a cosmological constant, are stable in the $ \bar{\mu} $ scheme. This rules out the $ \mu_o $ scheme in favor of the $ \bar{\mu} $ scheme for this particular quantization. The $\bar \mu$ scheme in mLQC-I resolved the problems at large volumes of the $\mu_o$ scheme in a very similar way as in LQC.

\section{von Neumann stability analysis of \lowercase{m}LQC-II quantization} \label{mLQC-II}

In mLQC-II, the departure from standard LQC arises in using eq. \eqref{KA} to express the extrinsic curvature in terms of the connection and obtain the expression \eqref{alternativeT} for the Lorentzian term. The quantization of the resulting Hamiltonian constraint leads to a quantum theory for the FLRW universe that yields qualitatively different physics than mLQC-I. An exploration of the phenomenological features in the mLQC-II in the $ \bar{\mu} $ scheme has been carried out in Ref. \cite{ps_mLQC-II}. A qualitative difference is that mLQC-II replaces the classical cosmological singularity by a symmetric bounce as in standard LQC, whereas the bounce is asymmetric in mLQC-I. Another important difference is that in mLQC-I one needs a non-trivial coupling between two branches in the gravitational phase space variable $b$ conjugate to volume $v$ to describe entire evolution, where as in mLQC-II one branch completely decouples and the physically relevant solution is described by only one of the branches \cite{ps_mLQC-II}. This feature will become crucial in the following discussion.  

We consider the von Neumann stability for the difference equations of mLQC-II obtained for both the $ \mu_o $ and $ \bar{\mu} $ schemes, where we note that the latter has been earlier obtained for this quantization in Ref. \cite{Yang2009}. 
We find that one of roots of the fourth order algebraic equation for the amplification factor is always larger than unity. Due to this one may be prompted to incorrectly conclude that this quantization is unstable for both $ \mu_o $ and $ \bar{\mu} $ scheme.
  Note that unlike the case of $\mu_o$ scheme in mLQC-I in presence of cosmological constant where one root in each pair of roots grew above unity, in the present case one of the pairs of roots is well behaved for the $\bar \mu$ scheme with a massless scalar field, and a cosmological constant, and in $\mu_o$ scheme with a massless scalar field. Thus, physical solutions constructed from the well behaved pair of roots would turn out to be stable in these three cases. This result is in synergy with the observations in effective dynamics of mLQC-II 
for the $\bar \mu$ scheme where no departures from GR have been found at large volumes for the physically relevant branch even in presence of inflationary potentials \cite{ps_mLQC-II}. As mentioned above, of the two mathematically allowed branches of $b$ only one branch contributes to the physical solutions and the other is completely irrelevant \cite{ps_mLQC-II}. If mLQC-II indeed was von Neumann unstable for the $\bar \mu$ scheme, it would have led to large departures such as a recollapse in the inflationary potentials for studies conducted in Ref. \cite{ps_mLQC-II}. These results from completely different starting points suggest that the physically relevant branch in the effective dynamics of mLQC-II is linked with the pair of roots bounded by unity, and the unphysical branch with the other pair.

\subsection{mLQC-II with scalar field in the $ \mu_o $ scheme} \label{mLQC-II-1}

To obtain the difference equation for mLQC-II, we only need to find out the operator for the Lorentzian term $ \mathring{T} $ given in \eqref{alternativeT}, as the expression for the Euclidean term, the corresponding operator and its action on the triad states remains the same as mLQC-I which have been evaluated in the previous section. The operator for the $ \mathring{T} $ is given by,
\begin{equation}
{\mathring{T}}^{\mu_o} = - \frac{24 i}{\kappa^2 \hbar \gamma^3} \sin\bigg(\frac{\mu_o c}{2}\bigg) \frac{\mathrm{sgn}(v)}{\mu_o^3} \bigg(\sin\bigg(\frac{\mu_o c}{2}\bigg) V \cos\bigg(\frac{\mu_o c}{2}\bigg) - \cos\bigg(\frac{\mu_o c}{2}\bigg) V \sin\bigg(\frac{\mu_o c}{2}\bigg)\bigg) \sin\bigg(\frac{\mu_o c}{2}\bigg).
\end{equation}

Its action on the triad basis can be easily calculated, and the resulting  difference equation for the massless scalar field becomes,
\ba
-\frac{\gamma^2}{2} W(3,1) |n+4\mu_o \rangle +2 (1+\gamma^2) W(2,0) |n+2\mu_o \rangle && \nonumber \\
+ \left\{ 2(1+\gamma^2) W(2,-2) - \frac{\gamma^2}{2} (W(1,3)+W(-3,-1))  \right\} |n\rangle && \nonumber \\
+2 (1+\gamma^2) W(0,-2) |n-2\mu_o \rangle - \frac{\gamma^2}{2} W(-1,-3) |n-4\mu_o \rangle && \nonumber \\
- \frac{\kappa^2 \hbar \gamma^3 \mu_o^3 B(n)}{6} \bigg(\frac{8\pi \gamma \lp^2}{6} \bigg)^{-3/2} \partial_{\phi}^2 |n\rangle &=& 0 \label{DifferenceAmu0scalar}
\ea

For the von Neumann analysis, we use the ansatz,
\begin{equation}\label{key}
|n\rangle = g^{n/2\mu_o} e^{i\omega \phi},
\end{equation}
which, at large volumes, reduces the above difference equation to,
\begin{equation}
g^4 - 16 R g^3 - 2(1-16R)g^2 - 16Rg + 1 = 0 ~.
\end{equation}
The roots of this equation come in two pairs:
\begin{equation}\label{rootsAmu0}
1,1; ~ \frac{2+\gamma^2 -2\sqrt{1+\gamma^2}}{\gamma^2},\frac{2+\gamma^2 +2\sqrt{1+\gamma^2}}{\gamma^2}.
\end{equation}
One pair consists of identical roots equal to unity. The second root from the other pair is clearly greater than unity. As discussed above, one should be careful in concluding that above behavior of roots signals a von Neumann instability. If the physical solutions are constructed from the pair of roots which are bounded by unity, then such solutions would not be problematic. This is confirmed by the phenomenological analysis of solutions in effective dynamics which show that the bounce is symmetric as in standard LQC \cite{ps_mLQC-II}. We also note that in standard LQC, the roots for the corresponding case are just one pair which equal unity. This seems to indicate that the difference equation here has a sector which is similar in behavior to the standard LQC dynamics. 

\subsection{mLQC-II with scalar field and a positive cosmological constant in the $ \mu_o $ scheme} \label{mLQC-II-2}

\begin{figure}
	\includegraphics[width=.4\textwidth]{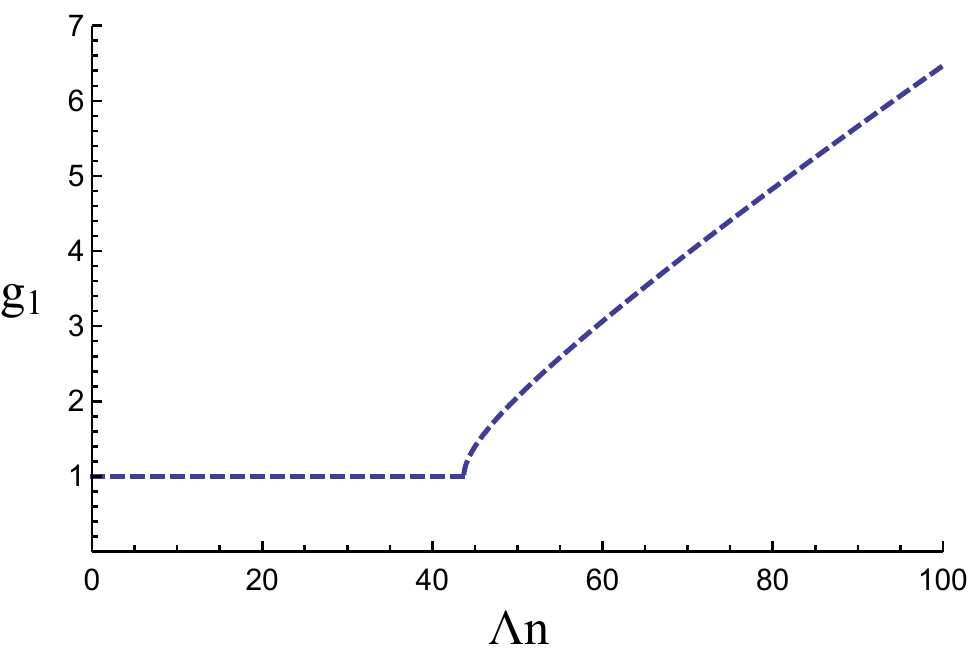}
	\includegraphics[width=.4\textwidth]{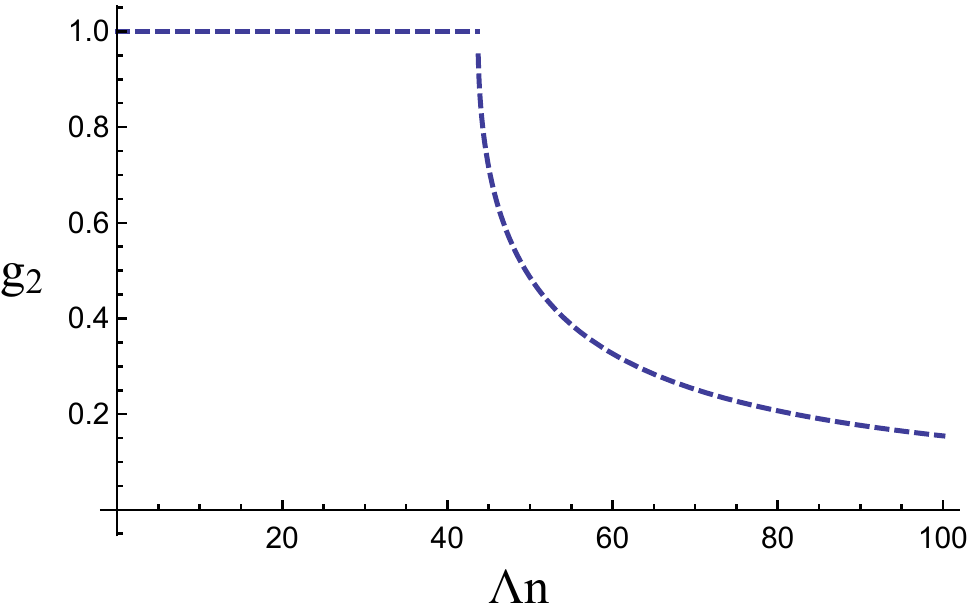}
	\includegraphics[width=.4\textwidth]{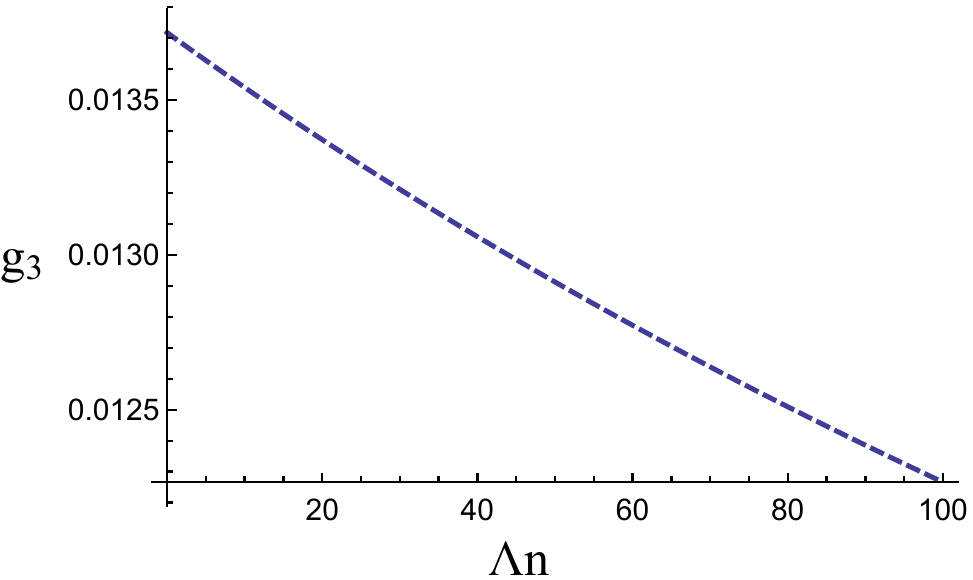}
	\includegraphics[width=.4\textwidth]{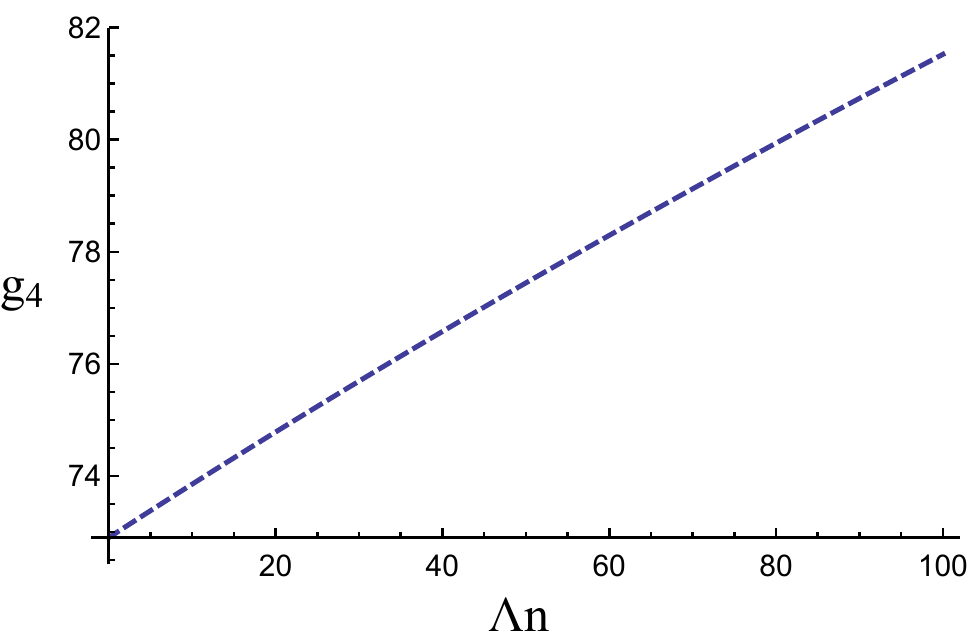}
	\caption{Absolute values of the roots are plotted for mLQC-II with scalar field and a positive cosmological constant in the $ \mu_o $ scheme.}
	\label{fig:fig3}
\end{figure}

Adding a positive cosmological constant changes the matter Hamiltonian as shown in equation \eqref{Hmatter_cosmologicalconstant}. The difference equation \eqref{DifferenceAmu0scalar} gets modified to,
\ba
-\frac{\gamma^2}{2} W(3,1) |n+4\mu_o \rangle +2 (1+\gamma^2) W(2,0) |n+2\mu_o \rangle && \nonumber \\
+ \left\{ 2(1+\gamma^2) W(2,-2) - \frac{\gamma^2}{2} (W(1,3)+W(-3,-1))  \right\} |n\rangle && \nonumber \\
+2 (1+\gamma^2) W(0,-2) |n-2\mu_o \rangle - \frac{\gamma^2}{2} W(-1,-3) |n-4\mu_o \rangle && \nonumber \\
- \frac{\kappa^2 \hbar \gamma^3 \mu_o^3 B(n)}{6} \bigg(\frac{8\pi \gamma \lp^2}{6} \bigg)^{-3/2} \partial_{\phi}^2 |n\rangle &=& \frac{\mu_o^3 \kappa \hbar \gamma^3}{3} \Lambda n^{3/2} |n\rangle, \label{DifferenceAmu0lambda}
\ea
which leads to the following fourth order polynomial equation for the amplification factor:
\begin{equation}
g^4 - 16 R g^3 - 2\bigg(1-16R + \frac{\mu_o^2 \kappa \hbar \gamma}{9} \Lambda n \bigg)g^2 - 16Rg + 1 = 0,
\end{equation}
The roots of the above equation in Planck units are,
\ba
g_{1,2} &=& \frac{1}{\gamma^2}[1-\sqrt{1+R_2 \Lambda n} + \gamma^2 \mp \sqrt{R_2 \Lambda n - 2(1+\gamma^2) (-1+\sqrt{1+R_2 \Lambda n})} ] \\
g_{3,4} &=& \frac{1}{\gamma^2}[1+\sqrt{1+R_2 \Lambda n} + \gamma^2 \mp \sqrt{R_2 \Lambda n + 2(1+\gamma^2) (1+\sqrt{1+R_2 \Lambda n})} ]
\ea
where $ R_2 := 16 \pi^2 \gamma^6 /3\sqrt{3} $. In the limit of vanishing $ \Lambda $, the pair $ (g_1,g_2) $ reduces to the pair $ (1,1) $ of the roots in \eqref{rootsAmu0} of the equation without cosmological constant. The pair $ (g_3,g_4) $ corresponds to the other pair of roots in \eqref{rootsAmu0}. As the roots depend on the combination $ \Lambda n $, we plot their magnitudes as a function of this quantity in Fig. \ref{fig:fig3}.
Unlike the case in Sec. IVA, now both the pair of roots are problematic for any given choice of cosmological constant. Thus, there is no suitable pair of roots which can allow a physically viable solution. At the level of effective dynamics, this means that even the branch which results in a physical solution will be problematic because of large departures from GR at large volumes. Thus, in the presence of a positive cosmological constant the $\mu_o$ quantization in mLQC-II suffers the same fate as in LQC and mLQC-I. It is von Neumann unstable and physically non-viable.

\subsection{mLQC-II with scalar field in the $ \bar{\mu} $ scheme} \label{mLQC-II-3}

The difference equation for mLQC-II in $ \bar{\mu} $ scheme has already been obtained in \cite{Yang2009} for lapse $ N=1 $. Since the expression for the operator of the Euclidean part remains the same as in \eqref{Euclidean_barmu}, we give here the operator for the Lorentzian part,
\begin{equation}
{\mathring{T}}^{\bar{\mu}} = - \frac{24 i}{\kappa^2 \hbar \gamma^3} \sin\bigg(\frac{\bar{\mu} c}{2}\bigg) \frac{\mathrm{sgn}(v)}{\bar{\mu}^3} \bigg(\sin\bigg(\frac{\bar{\mu} c}{2}\bigg) V \cos\bigg(\frac{\bar{\mu} c}{2}\bigg) - \cos\bigg(\frac{\bar{\mu} c}{2}\bigg) V \sin\bigg(\frac{\bar{\mu} c}{2}\bigg)\bigg)\sin \bigg(\frac{\bar{\mu} c}{2}\bigg).
\end{equation}

The quantum difference equation obtained in this case in the volume representation is \cite{Yang2009},
\begin{eqnarray}\label{key}
\frac{C_2}{2} |v| \bigg||v+1|^{1/3} - |v-1|^{1/3} \bigg|^3 \partial_{\phi}^2 |v\rangle &=& f'_{+} (v)|v+4\rangle + S'_{+} (v)|v+2\rangle \nonumber\\
&& + (f'_o (v) + S'_o (v))|v\rangle \nonumber\\
&& + S'_{-} (v)|v-2\rangle + f'_{-} (v)|v-4\rangle,
\end{eqnarray}

where $ f'_{+}(v), f'_{-}(v) $ and $ f'_o (v) $ have already been defined above. The rest of the coefficients are given by,
\begin{eqnarray}
S'_{+} (v) &=& 16R C_3 (v+1) (|v+2|-|v|), \\
S'_{-} (v) &=& S'_{+} (v-2), \\
S'_o (v) &=& - S'_{+} (v) - S'_{-} (v).
\end{eqnarray}

Using the following ansatz to do the von Neumann stability analysis,
\begin{equation}\label{key}
|v\rangle = g^{\frac{v}{2}} e^{i\omega \phi},   
\end{equation}
we obtain the fourth order polynomial equation in $ g $:
\begin{equation}\label{key}
-\frac{C_2}{2} |v| \bigg||v+1|^{1/3} - |v-1|^{1/3} \bigg|^3 \omega^2 g^2 = f'_{+} (v)g^4 + S'_{+} (v)g^3 + (f'_o (v) + S'_o (v))g^2 + S'_{-} (v)g + f'_{-} (v).
\end{equation}
In the limit of large volume, this reduces to the equation,
\begin{equation}\label{key}
g^4 - 16R g^3 + 2(16R-1)g^2 - 16Rg +1 =0.
\end{equation}
This is the same as the equation obtained for mLQC-II in the $ \mu_o $ scheme in case of a massless scalar field in Sec. IVA. As already pointed out, we have two pairs of roots given in \eqref{rootsAmu0}, and only the second pair has one root that is larger than unity, while the first pair has roots equal to unity. As in $\mu_o$ scheme, if physical solutions can be constructed using the well behaved pair of roots then stability is guaranteed which is consistent with the analysis of effective dynamics in Ref. \cite{ps_mLQC-II}.

\begin{figure}
	\includegraphics[width=.4\textwidth]{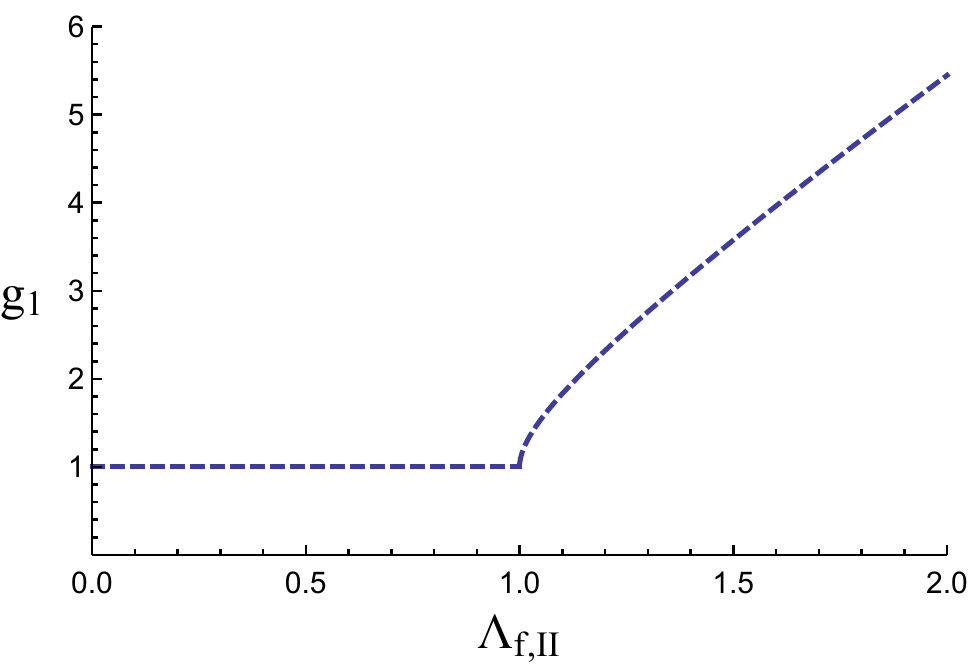}
	\includegraphics[width=.4\textwidth]{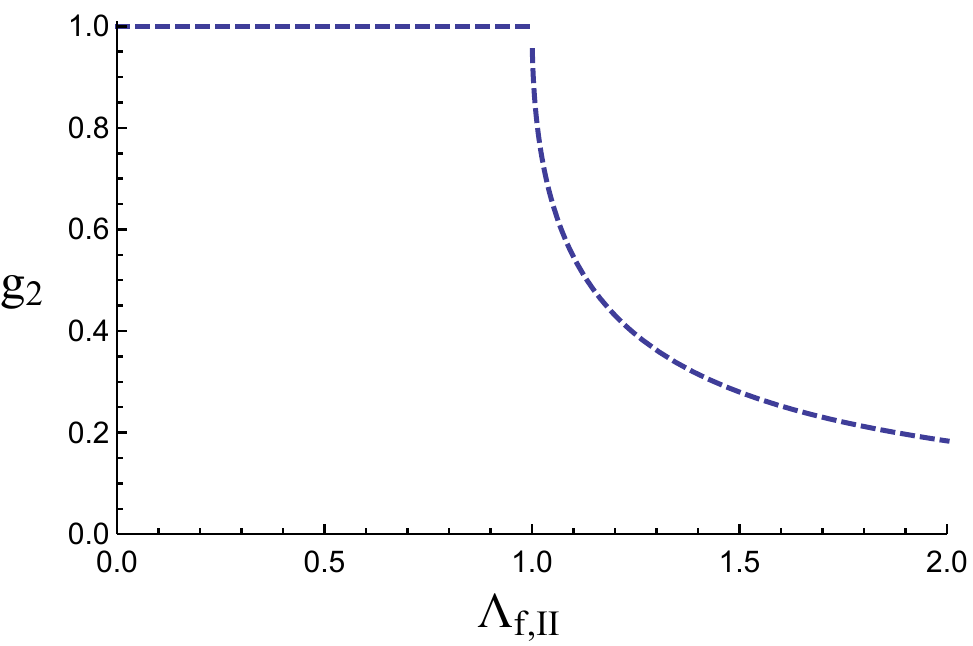}
	\includegraphics[width=.4\textwidth]{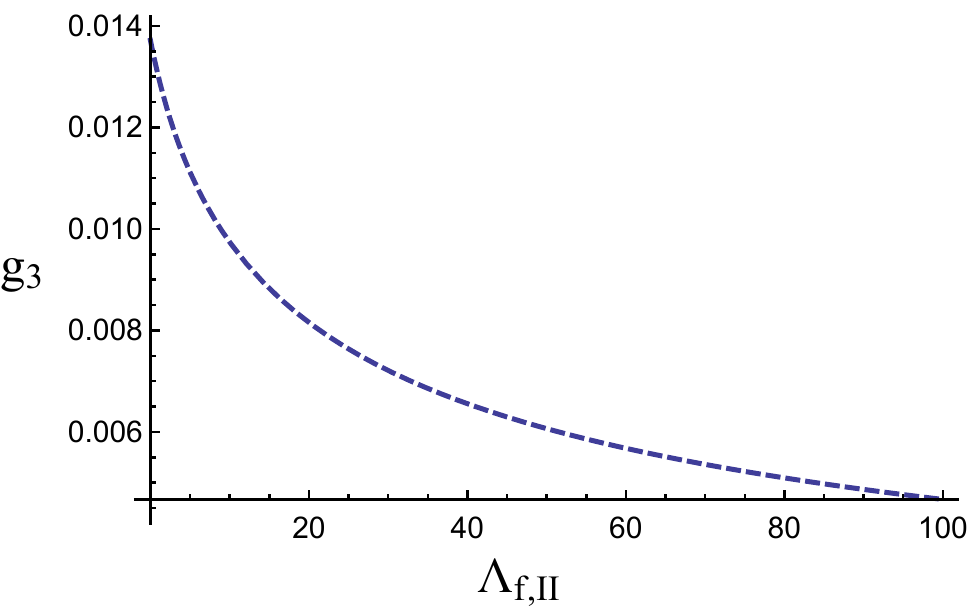}
	\includegraphics[width=.4\textwidth]{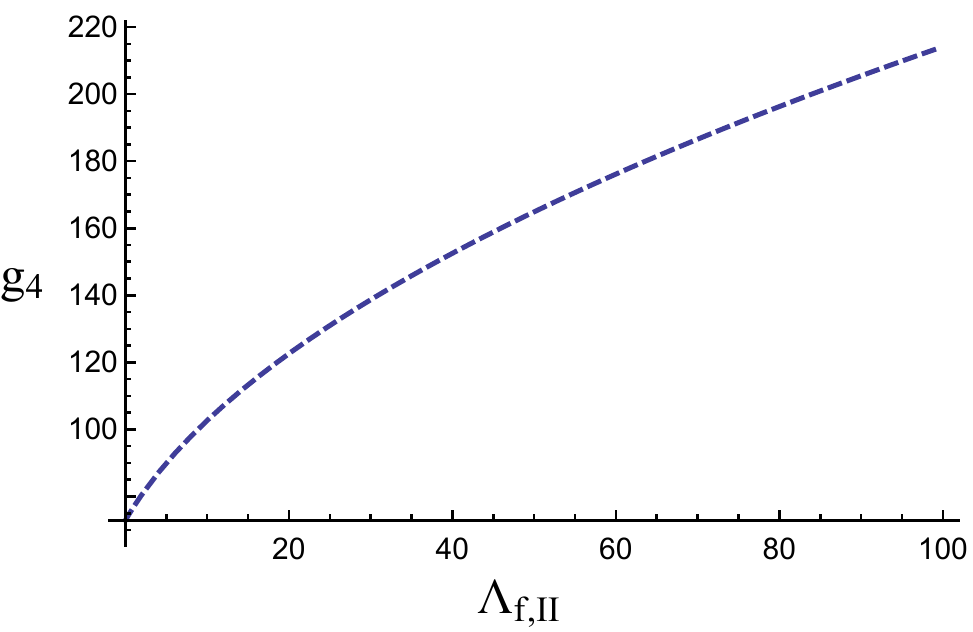}
	\caption{Absolute values of the roots are plotted for mLQC-II with scalar field and a positive cosmological constant in the $ \bar{\mu} $ scheme.}
	\label{fig:fig4}
\end{figure}

\subsection{mLQC-II with scalar field and a positive cosmological constant in the $ \bar{\mu} $ scheme} \label{mLQC-IV-4}
For the $\mu_0$ quantization in mLQC-II, adding a positive cosmological constant resulted in von Neumann instability at large volumes. We now see  the consequences in $\bar \mu$ scheme, where adding a positive cosmological constant in the quantum Hamiltonian constraint  results in the following difference equation:
\begin{eqnarray}\label{key}
\frac{C_2}{2} |v| \bigg||v+1|^{1/3} - |v-1|^{1/3} \bigg|^3 \partial_{\phi}^2 |v\rangle + \frac{\gamma \lp^2 \sqrt{\Delta}}{4} \Lambda v |v\rangle &=& f'_{+} (v)|v+4\rangle + S'_{+} (v)|v+2\rangle \nonumber\\
&& + (f'_o (v) + S'_o (v))|v\rangle + S'_{-} (v)|v-2\rangle \nonumber\\
&& + f'_{-} (v)|v-4\rangle .
\end{eqnarray} 
This leads to the following equation for the amplification factor $ g $ in the large volume limit:
\begin{equation}\label{key}
g^4 - 16R g^3 + \bigg(2(16R-1) - 64R  \Lambda_{f,\mathrm{II}} \bigg) g^2 - 16 R g + 1 = 0,
\end{equation}
where  $ \Lambda_{f,\mathrm{II}}:=\Lambda /\Lambda_{c,\mathrm{II}} $, and $\Lambda_{c,\mathrm{II}} $ is the critical value of the cosmological constant which is equal to $ \kappa $ times the critical density in mLQC-II. The critical density in this case is $ 3(\gamma^2 +1)/(2\pi G \gamma^2 \Delta) $, which is the maximum energy density attainable as deduced from the effective dynamics of this model \cite{Yang2009,ps_mLQC-II}. The above equation has four roots,
\ba
g_{1,2} &=& 1+\frac{1}{\gamma^2} \bigg[1-\sqrt{1+4\gamma^2(1+\gamma^2)\Lambda_{f,\mathrm{II}}} \nonumber \\
&& \pm \sqrt{2(1+\gamma^2)\bigg(1+2\gamma^2\Lambda_{f,\mathrm{II}} - \sqrt{1+4\gamma^2(1+\gamma^2)\Lambda_{f,\mathrm{II}}}\bigg)}\bigg] \\
g_{3,4} &=& 1+\frac{1}{\gamma^2}\bigg[1+\sqrt{1+4\gamma^2(1+\gamma^2)\Lambda_{f,\mathrm{II}}} \nonumber \\
&& \pm \sqrt{2(1+\gamma^2)\bigg(1+2\gamma^2\Lambda_{f,\mathrm{II}} + \sqrt{1+4\gamma^2(1+\gamma^2)\Lambda_{f,\mathrm{II}}}\bigg)}\bigg] 
\ea
The pair $ (g_1,g_2) $ reduce to $ 1 $ in the limit $ \Lambda \rightarrow 0 $. While the pair $ (g_3,g_4) $ reduce to the second pair in \eqref{rootsAmu0} in the limit $ \Lambda \rightarrow 0 $. Since the maximum value of the cosmological constant allowed is $ \Lambda = \Lambda_{c,\mathrm{II}} $, we expect the difference equation to be unstable for $ \Lambda_{f,\mathrm{II}} > 1 $, which is indeed the case as is evident from Fig. \ref{fig:fig4} where we plot the roots as a function of $ \Lambda_{f,\mathrm{II}} $.

We note that the second pair of roots, which was problematic even in absence of cosmological constant and belonged to the unphysical branch of evolution, remains problematic. For this pair, one of the roots, $ g_4 $ is larger than unity for all values of $ \Lambda_{f,\mathrm{II}} $. On the other hand, the first pair of roots has magnitude unity for $ \Lambda_{f,\mathrm{II}} \in [0,1] $. If physical solutions are constructed from this pair of roots then stability is obtained, which is   consistent with results from effective dynamics on inflationary potentials \cite{ps_mLQC-II}. Recall that the difference equation was unstable in the $ \mu_o $ scheme when we included a positive cosmological constant in mLQC-II. In that case, both the pair of roots were problematic and there would be no physically viable solution. This rules out the $ \mu_o $ scheme for mLQC-II and indicates that the $ \bar{\mu} $ scheme is the right choice for mLQC-II quantization.

\section{Summary}
The von Neumann stability analysis has proved to be quite useful in gaining insights on the viability of loop quantum cosmological models at infra-red scales. Quantization of Hamiltonian constraint in these models result in a quantum difference equation which is expected to approximate the differential Wheeler-DeWitt equation at large volumes. Since the solutions of Wheeler-DeWitt equation at large volume capture those of GR, an agreement between solutions of loop quantum difference equations and Wheeler-DeWitt equation ensures that GR is recovered in large volume limit from loop quantum cosmological models. As an example, the $\mu_o$ scheme in LQC (where  $\mu_o$ is a numerical constant) or old LQC \cite{Ashtekar2003,aps2}  is von Neumann unstable in presence of positive cosmological constant \cite{ps12}. This instability reveals itself at the level of effective dynamics in terms of an unphysical recollapse of an inflating universe at large volumes \cite{cs-unique}, and in undesirable properties of eigenfunctions \cite{noui}. On the other hand, for the $\bar \mu$ scheme (or the improved dynamics), the quantum difference equation is stable even in presence of positive cosmological constant till the corresponding energy density reaches the maximal allowed value in Planck regime \cite{tanaka,ps12}.  The problems with von Neumann stability of quantum difference equation and undesirable phenomenological implications are inter-linked, presence of one often signaling existence of the  other.

In this manuscript, we performed the von Neumann stability analysis of the difference equations in mLQC-I and mLQC-II quantizations of the spatially flat FLRW model in the $ \mu_o $ and $ \bar{\mu} $ schemes. The mLQC-I and mLQC-II quantizations result from quantization ambiguities in obtaining quantum Hamiltonian constraint which yield qualitatively different physics from standard LQC. In particular, the quantum Hamiltonian constraint in both of these quantizations results in a finite difference equation which is of fourth order, unlike the second order equation in LQC. Using von Neumann stability analysis a fourth order polynomial equation for the amplification factor is found at large volume which results in two pairs of roots. Analyzing $\mu_o$ and $\bar \mu$ schemes for mLQC-I and mLQC-II quantizations we explored which of the choices leads to a viable prescription for matter content which is massless scalar field, and massless scalar field with a positive cosmological constant. Such a choice of matter ensures that the entire range of equation of state from positive unity (massless scalar) to negative unity (cosmological constant) is explored. Thus, one can reach conclusions for stability of the quantum difference equation for matter satisfying weak energy condition.

We found that the difference equation in mLQC-I for the $ \mu_o $ scheme is von Neumann stable if we have a scalar field as matter. However, it is unstable if we also include a positive cosmological constant in the matter Hamiltonian as two of the roots, one from each pair, exceed unity. On the other hand,  the difference equation for mLQC-I is stable in the $ \bar{\mu} $ scheme even after including a positive cosmological constant in the matter Hamiltonian as none of the four roots exceed unity in this case. The effective dynamics of mLQC-I for inflationary potentials has been recently studied in detail in Ref. \cite{ps_mLQC-II}, and the results found here are in synergy. 

Unlike mLQC-I, we found that in mLQC-II, one of the roots from a pair always exceeds unity. However, one pair yielded roots which equaled unity for the case of $\bar \mu$ scheme with a massless scalar field, and with a cosmological constant, and $\mu_o$ scheme with a massless scalar field. Physical solutions  constructed from this pair of  roots would reliably capture classical solutions at large volumes. Here  phenomenological insights obtained from the analysis of effective dynamics of mLQC-II model \cite{ps_mLQC-II} turn out to be very useful. In the latter study detailed investigation of mLQC-II effective dynamics with inflationary potentials reveals no departures from GR at large volume or an unphysical recollapse which is a hallmark of associated von Neumann instability. Thus, care must be exercised in concluding about von Neumann instability from just one pair of roots. We noted that the phase space of mLQC-II is divided into two branches each having its own Friedmann and Raychaudhuri equations. While in mLQC-I, both the sets of Friedmann and Raychaudhuri equations (i.e. both branches) are needed to describe the complete phase space trajectory, the two branches decouple in mLQC-II and only one of the branches corresponds to the complete physical evolution in case of mLQC-II. The other branch leads to unphysical evolution in mLQC-II. These results indicate that the physically relevant branch corresponds to the pair of roots which equal unity in von Neumann stability analysis. For the case of  massless scalar field, we found that mLQC-II is von Neumann stable in $ \mu_o $ scheme. However, in $ \mu_o $ scheme, the quantum difference equation becomes unstable when a positive cosmological constant is included in the matter Hamiltonian. This is because in this case both the pairs of roots have a root which grows greater than unity at some large volume. Thus, no physically viable solution is possible for the $\mu_o$ scheme.  As in the case of mLQC-I, the $ \bar{\mu} $ scheme turned out to be stable in absence and presence of positive cosmological constant for mLQC-II.

In summary, using the von Neumann stability analysis we found out that the $ \mu_o $ scheme is unstable in both mLQC-I and mLQC-II when cosmological constant is included. Thus, the $ \mu_o $ scheme is phenomenologically ruled out for modified LQC, as was the case for standard LQC. However, the $ \bar{\mu} $ scheme  gives stable difference equations for both mLQC-I and mLQC-II in presence of cosmological constant, pointing to necessity as well as robustness of improved dynamics, first introduced in Ref. \cite{aps3}, for modified loop quantum cosmologies.

\section*{Acknowledgments}
 This work is supported by  NSF grant PHY-1454832.
 
 \begin{appendix}
\section{Agreement between loop quantum difference equations and Wheeler-DeWitt differential equations at large volumes} \label{a1}

At large volumes where $n$ is very large compared to unity, the quantum discreteness scale becomes negligible. In this regime, one can show that for semi-classical wavefunctions the Wheeler-DeWitt differential equation approximates the LQC quantum difference equation (see for eg. \cite{Bojowald2002,Ashtekar2003,ps12}). 
In this appendix, we give a brief summary of this result for the case of mLQC-I for the $\mu_o$  with massless scalar field as matter. 
A similar analysis holds for the $\bar \mu$ scheme in mLQC-I, and for $\mu_o$ and $\bar \mu$ prescription in mLQC-II. 

For $\mu_o$ prescription of mLQC-I discussed in Sec. IIIA, the operator form of the Euclidean part of the Hamiltonian constraint is given by,
	\begin{equation}\label{euclidean_mu_o}
	H^{E,\mu_o} = - \frac{24 \gamma^2  i}{2 \kappa^2 \hbar \gamma^3} \sin(\mu_o c) \frac{\mathrm{sgn}(v)}{\mu_o^3}\bigg(\sin\bigg(\frac{\mu_o c}{2}\bigg) V \cos\bigg(\frac{\mu_o c}{2}\bigg) - \cos\bigg(\frac{\mu_o c}{2}\bigg) V \sin\bigg(\frac{\mu_o c}{2}\bigg)\bigg) \sin(\mu_o c) ~
	\end{equation}
where we have ignored hats on operators for brevity. 	
To simplify our notation let us choose $\mu_o = 1$ and the orientation of the triad to be positive. From eq.(\ref{triadev}), the eigenvalues of the triad operator $\hat p$ are related to the $n$ labels of the quantum state as $p= \sigma n $ where $ \sigma:=\kappa \hbar \gamma /6 $. We are interested in the large volume regime, i.e. when $n \gg 1$, and the change between $n$ and $n+1$ is negligible for a classical observer.  We consider smooth wavefunctions $\psi_n(p)$ which interpolate between discrete labels specified by $n$ at large volumes.\footnote{Behavior of such wavefunctions for massless scalar field matter is discussed in Refs. \cite{aps2,aps3}.}.  For such wavefunctions, we show that the quantum difference operators in LQC can be well approximated by the differential operators in Wheeler-DeWitt theory in the large volume regime.


	
	Let us consider the operator $S := 2i \sin(c/2) $ at large volumes:
	\begin{equation}\label{key}
	S \psi_n = \psi_{n+1}-\psi_{n-1}  \approx \frac{\d\psi_n}{\d n}\delta n
	\end{equation}
	where we have ignored higher order terms in the Taylor approximation.  Using $ \delta n = (n+1)-(n-1) = 2 $, and the relation between triads and labels $n$ we get 
		\begin{equation}\label{key}
	S \psi_n \approx 2\sigma \frac{\d\psi_n}{\d p} .
	\end{equation}
		Thus, $S$ becomes the above mentioned differential operator under the semi-classical approximation at large volumes. Similarly, the action of the operator $C := \cos(c/2) $ on the smooth semi-classical wavefunction is just the average of the wave function at $ (n-1) $ and $ (n+1) $ to the leading order at large volumes, which  is well approximated by $ \psi_n $. 
	
	Further, one can show that,
	\begin{equation}\label{key}
	\bigg(\sin\bigg(\frac{\mu_o c}{2}\bigg) V \cos\bigg(\frac{\mu_o c}{2}\bigg) - \cos\bigg(\frac{\mu_o c}{2}\bigg) V \sin\bigg(\frac{\mu_o c}{2}\bigg)\bigg) \psi_n = \frac{i \sigma^{3/2}}{2} \bigg(|n+1|^{3/2} - |n-1|^{3/2} \bigg) \psi_n
	\end{equation}
	can be approximated at large volumes as:
	\begin{equation}\label{key}
	\bigg(\sin\bigg(\frac{c}{2}\bigg) V \cos\bigg(\frac{ c}{2}\bigg) - \cos\bigg(\frac{ c}{2}\bigg) V \sin\bigg(\frac{ c}{2}\bigg)\bigg) \psi_n \approx i\frac{3 \sigma}{2} p^{1/2} \psi_n
	\end{equation}
		
	The Euclidean term in the Hamiltonian constraint, can thus be approximated by the following differential operator at large volumes, 

	\begin{eqnarray}\label{key}
	H^{E} \psi_n && = - \frac{12}{\kappa^2 \hbar \gamma} S C \bigg(\sin\bigg(\frac{ c}{2}\bigg) V \cos\bigg(\frac{ c}{2}\bigg) - \cos\bigg(\frac{ c}{2}\bigg) V \sin\bigg(\frac{ c}{2}\bigg)\bigg) C S \psi_n \\
	&& \approx - \frac{72}{\kappa^2 \hbar \gamma} \sigma^3 \frac{\d}{\d p} p^{1/2} \frac{\d}{\d p} \psi_n .\\
	\end{eqnarray}
	Thus we obtain the approximation of the Euclidean part of the Hamiltonian constraint in the semi-classical regime in terms of differential operators. Noting that in the $p$ representation the action of $\hat c$ is given by $\hat c \psi = i \hbar \frac{\kappa \gamma}{3} \frac{\d}{\d p} \psi$, we find  
	\be
	H^{E} \psi_n= \frac{3}{\kappa} {c} \, {{p^{1/2}}} \, {c} \psi_n.
	\ee
	Comparing with eq.(2.4), we find that the Euclidean part of the quantum constraint in loop quantum cosmologies turns out to be same as the  Wheeler-DeWitt differential operator for the Euclidean part at large volumes. 
	
	The Lorentzian part for the Hamiltonian constraint in mLQC-I is as follows:
	\ba
	T^{\mu_o} &=& - \frac{24 i}{\kappa^4 \hbar^5 \gamma^7} \bigg(\sin\bigg(\frac{\mu_o c}{2}\bigg) B \cos\bigg(\frac{\mu_o c}{2}\bigg) - \cos\bigg(\frac{\mu_o c}{2}\bigg) B \sin\bigg(\frac{\mu_o c}{2}\bigg)\bigg) \nonumber \\
	&& \times \frac{\mathrm{sgn}(v)}{\mu_o^3}\bigg(\sin\bigg(\frac{\mu_o c}{2}\bigg) V \cos\bigg(\frac{\mu_o c}{2}\bigg) - \cos\bigg(\frac{\mu_o c}{2}\bigg) V \sin\bigg(\frac{\mu_o c}{2}\bigg)\bigg) \nonumber \\
	&& \times \bigg(\sin\bigg(\frac{\mu_o c}{2}\bigg) B \cos\bigg(\frac{\mu_o c}{2}\bigg) - \cos\bigg(\frac{\mu_o c}{2}\bigg) B \sin\bigg(\frac{\mu_o c}{2}\bigg)\bigg). \label{Lorentzian_mu_o}
	\ea
	Here we have defined the operator $ B = [H^{E,\mu_o}, V] $. Using similar arguments as above, the action of the following operator can be approximated in the large volume limit as:
	
	\begin{equation}\label{key}
	\bigg(\sin\bigg(\frac{c}{2}\bigg) B \cos\bigg(\frac{ c}{2}\bigg) - \cos\bigg(\frac{ c}{2}\bigg) B \sin\bigg(\frac{ c}{2}\bigg)\bigg) \psi_n \approx i \frac{108 \sigma^3}{\kappa^2 \hbar \gamma} S \psi_n.
	\end{equation}
	
	Then the Lorentzian term corresponds to the following differential operator in the large volume limit:	
	\begin{eqnarray}\label{key}
	T && \approx - \frac{3}{2 \kappa \gamma^2}(4\sigma^2) \frac{\d}{\d p} p^{1/2}  \frac{\d}{\d p} \psi_n\\
	&& = \frac{3}{2 \kappa \gamma^2} {c} {{p^{1/2}}} {c} \psi_n. 
	\end{eqnarray}
	which is the same as the Wheeler-DeWitt operator for the classically reduced expression for the
Lorentzian part given in eq. (2.5).
	
	
	Hence, the Euclidean and Lorentzian parts of the Hamiltonian constraint in mLQC-I for $\mu_o$ quantization with a massless scalar field are well approximated by the corresponding Wheeler-DeWitt operators in the semi-classical regime at large volumes. A similar analysis for mLQC-II yields the same result.

\end{appendix}
  

\begin{thebibliography}{10}

 \bibitem{as-status}
	A. Ashtekar and P. Singh, Class. Quant. Grav. \textbf{28}, 213001, (2011).

\bibitem{aps3} A.~Ashtekar, T.~Pawlowski, and P.~Singh, Phys. Rev. D \textbf{74}, 084003, (2006).

\bibitem{aps2}
A.~Ashtekar, T.~Pawlowski, and P.~Singh, Phys. Rev. D \textbf{73},
 124038, (2006).

\bibitem{Bojowald2003}
M.~Bojowald and G.~Date,  Class. Quant. Grav. {\bf 21}, 121, (2004).

\bibitem{date2}  G. Date,  Phys. Rev. D {\bf 72}, 067301, (2005).

\bibitem{Cartin2005}
D.~Cartin and G.~Khanna, Phys. Rev. Lett. \textbf{94}, 111302, (2005).

\bibitem{Cartin2005a}
D.~Cartin and G.~Khanna, Phys. Rev. D \textbf{72}, 084008, (2005).

\bibitem{Rosen2006}
J.~Rosen, J.-H. Jung, and G.~Khanna, Class. Quant. Grav. \textbf{23}, 7075,
(2006).

\bibitem{martin-cartin-khanna}
M.~Bojowald, D.~Cartin, and G.~Khanna, Phys. Rev. D \textbf{76}, 064018, (2007).

\bibitem{Nelson2009}
W.~Nelson and M.~Sakellariadou, Phys. Rev. D \textbf{80}, 063521, (2009).

\bibitem{tanaka} T. Tanaka \textit{et al.},  Phys. Rev. D {\bf 83}, 104049, (2011).

\bibitem{khanna-review}
D.~Brizuela, D.~Cartin, and G.~Khanna, SIGMA \textbf{8}, 001, (2012).

\bibitem{ps12}
P.~Singh, Class. Quant. Grav. \textbf{29}, 244002, (2012).

\bibitem{Cartin2006}
D.~Cartin and G.~Khanna, Phys. Rev. D \textbf{73}, 104009, (2006).

\bibitem{Yonika2018}
A.~Yonika, G.~Khanna, and P.~Singh, Class. Quant. Grav. \textbf{35}, 045007, (2018).

\bibitem{cs-unique}
A.~Corichi and P.~Singh, Phys. Rev. D \textbf{78}, 024034, (2008).

\bibitem{Yang2009}
J.~Yang, Y.~Ding, and Y.~Ma, Phys. Lett. B \textbf{682}, 1, (2009).

\bibitem{Dapor2017}
A.~Dapor and K.~Liegener, Phys. Lett. B \textbf{785}, 506, (2018).

\bibitem{tomasz} M.~Assanioussi, A.~Dapor, K.~Liegener and T.~Pawlowski,
Phys. Rev. Lett. {\bf 121}, 081303, (2018).

\bibitem{Li2018}
B.-F. Li, P.~Singh, and A.~Wang, Phys. Rev. D \textbf{97}, 084029, (2018).



\bibitem{ps_mLQC-II}
B.-F. Li, P.~Singh, and A.~Wang, Phys. Rev. D \textbf{98}, 066016, (2018).

\bibitem{Ashtekar2003}
A.~Ashtekar, M.~Bojowald, J.~Lewandowski, {\em et~al.}, Adv. in Theor. and
	Math. Phys. \textbf{7}, 233, (2003).



\bibitem{aps1}
A.~Ashtekar, T.~Pawlowski, and P.~Singh, Phys. Rev. Lett. \textbf{96}, 141301, (2006).

\bibitem{thiemann}  T.  Thiemann,  Class.  Quant.  Grav. {\bf 15},  839,  (1998);  T. Thiemann, Class. Quant. Grav. {\bf 15}, 875, (1998); K. Giesel, T. Thiemann, Class. Quant. Grav. {\bf 24}, 2465, (2007).


\bibitem{saini-singh-mlqc}  S.~Saini and P.~Singh, Class. Quantum Grav. {\bf 36}, 105014, (2019).






\bibitem{cs-geom} A.~Corichi and P.~Singh,
Phys.\ Rev.\ D {\bf 80}, 044024, (2009).


\bibitem{noui} K. Noui, A. Perez and K. Vandersloot,  Phys. Rev. D {\bf 71}, 044025, (2005).

\bibitem{agullo-singh} I. Agullo, P. Singh, {\it{Loop Quantum Cosmology}} in
{\it{Loop Quantum Gravity:  The First 30 Years}},  Eds:   A.  Ashtekar, J.   Pullin,   World   Scientific, (2017).

\bibitem{Bojowald2002}
M.~Bojowald, Class. Quant. Grav. \textbf{19}, 2717, (2002).




\bibitem{Strikwerda2004}
J.~C. Strikwerda, {\em Finite difference schemes and partial differential
	equations}, \newblock Siam, 2004.
	


\bibitem{Lax1956}
P.~D. Lax and R.~D. Richtmyer, Comm. Pure Appl. Math. \textbf{9}, 267, (1956).




  







  
  
\end{thebibliography}


\end{document}